\documentclass[amsmath,amssymb,aps,prd,11pt,tightenlines,superscriptaddress, 
nofootinbib,preprintnumbers,notitlepage]{revtex4-2}
\usepackage{graphicx}
\usepackage{subcaption}
\usepackage{dcolumn}
\usepackage{bm}
\usepackage{amssymb}
\usepackage{amsmath}
\usepackage{mathrsfs}
\usepackage{graphicx,color} 
\usepackage[english]{babel}     
\usepackage{graphicx,wrapfig,lipsum}  
\usepackage{enumitem}

\usepackage{float}

\usepackage{multimedia}    
\usepackage[mathscr]{eucal}  
\usepackage{bm}              
\usepackage{amssymb}           
\usepackage{amsmath}      
\usepackage{mathrsfs}  
\usepackage[mathscr]{eucal}      
\usepackage[normalem]{ulem}
\usepackage{cleveref}

\begin{document}

\title{Ensembles of center vortices and chains: Insights from a natural lattice framework}

\author{David R. Junior}

\affiliation{
Instituto de F\'isica Te\'orica, Universidade Estadual Paulista and South-American Institute of Fundamental Research ICTP-SAIFR, Rua Dr. Bento Teobaldo Ferraz, 271 - Bloco II, 01140-070 São Paulo, SP, Brazil} 
\author{Luis E. Oxman}
\affiliation{
Instituto de F\'\i sica, Universidade Federal Fluminense, 24210-346 Niter\'oi, RJ, Brasil.} 

\date{\today}

\begin{abstract}     

A scenario to understand the asymptotic properties of confinement between quark probes, based on a 4D mixed ensemble of percolating center-vortex worldsurfaces and chains, was initially proposed by one of us in a non-Abelian setting. More recently, the same physics was reobtained by means of a Schr\"odinger wavefunctional peaked at Abelian-projected configurations, which deals with center-vortex lines and pointlike monopoles in  real space. In this work, we formulate the Abelian-projected ensemble and reassess the non-Abelian one within the Weingarten lattice representation for the sum over surfaces. In the phase where worldsurfaces are stabilized by contact interactions and percolate, lattice gauge fields emerge. This generalizes the Goldstone modes in an Abelian loop condensate to the case where non-Abelian degrees of freedom are present. In this language, the different natural matching properties of elementary 
center-vortex worldsurfaces and monopole worldlines can be easily characterized. In the lattice, the Abelian setting also implements the original idea that the mixed ensemble reconciles $N$-ality with the formation of a confining flux tube. In this picture, center vortices and chains explain why Abelian-projected variables capture this property at asymptotic distances while simultaneously allowing for a 'dual superconductor' description of the fundamental string. Common features, differences in the continuum, and perspectives will also be addressed. \

\end{abstract} 

\maketitle

\section{Introduction}

Pioneering research proposed center vortices as fundamental topological structures for understanding the infrared dynamics of the $SU(N)$ Yang-Mills vacuum and the mechanism of confinement \cite{Mandelstam1976,tHooft1978,tHooft1979,MackPetkova1979,NielsenOlesen1979}.  Another line of research was opened in Ref. \cite{superconduct-1}, where the confining properties of non-Abelian gauge theories were proposed to be dominated by Abelian variables. This framework identified monopoles as topological defects in Abelian-projected fields, whose proliferation signals a dual superconductor phase, confining quarks via flux tubes.
These groundbreaking ideas guided a plethora of studies exploring both scenarios.    In the lattice, ensembles of percolating center-vortex networks, which naturally accommodate the $N$-ality property observed in the  Wilson loop averages at asymptotic distances  \cite{n-ality}, have been shown to play a central role \cite{cv-1,cv-2,cv-3,cv-4,percolating-1,randomsu2,randomsu3,cv-5,cv-6,cv-7,cv-9,cv-10,cv-11,cv-8}.  More recently, the importance of center vortices in describing infrared physics even in the presence of dynamical fermions was established in Refs. \cite{potential-fermions,percolating-2}.  However, center-projected scenarios, where configurations carry only $Z(N)$ charges and are propagated along worldsurfaces, do not explain the formation of a flux tube nor the ensuing transverse fluctuations manifested as a Lüscher term.
 On the Abelian-projected side, a large amount of information was also obtained by means of a lattice implementation \cite{ab-proj-0}.  Abelian projection gets things right at asymptotic distances: it captures $N$-ality \cite{ab-proj-1,ab-proj-2} and the potential between quarks in the defining representation displays a perfect Abelian dominance \cite{ab-proj-3,suga2014}. Additionally, effective descriptions in the spirit of Polyakov’s compact QED \cite{polyakov} have been proposed to understand the flux tube in monopole-only scenarios \cite{mon-eff-1,mon-eff-2,mon-eff-3,mon-eff-4}. Nonetheless,  there is a problem when trying to associate Abelian dominance with monopole-only scenarios, as the latter cannot describe $N$-ality \cite{d-wl}. To address this situation, our work has  focused on identifying a possible mechanism that reconciles $N$-ality with the formation of a confining flux tube, within the framework of ensembles of percolating topological objects. Physically, $N$-ality manifests when the distance between quark probes in a general $SU(N)$ representation becomes sufficiently large. At such distances the confining flux tube observed at intermediate separations eventually transitions to the lowest-energy state with the same $N$-ality, due to the creation of valence gluons that screen the probes.
It is also in this regime that the observed center-vortex thickness \cite{latt-thickness} can be disregarded compared to the Wilson loop size, making a model based on the guiding-centers of center vortices feasible.    In Ref. \cite{mixed}, it was proposed that mixed ensembles of center vortices and chains could integrate the various asymptotic properties of confinement in $SU(N)$ YM theory. This proposal was motivated by the $N$-ality properties of the former and the possibility to trigger a flux tube with the monopole component in chains, where elementary center vortices (center-charge $\pm 1$) carrying different defining magnetic weights $\beta_i$, $i=1, \dots , N$, are attached in pairs to monopoles with charge $\beta_i - \beta_j$. Indeed, the Abelian-projected link-variables in Monte Carlo simulations of $SU(2)$ YM theory reveal these collimated fluxes \cite{collimation},  also known as nonoriented center vortices, which  contribute to the topological charge \cite{nonoriented-1}.  In $SU(2)$, when these variables were further processed to get center projected ones, it was established that $61 \%$ of the configurations contained no monopoles, while the rest contained at least a monopole-antimonopole pair. In Ref.~\cite{Stack2002}, similar properties were studied in $SU(3)$ YM theory. Although the  Cartan 
flux-collimation along center vortices  was not addressed,  monopoles and center vortices were  detected using Abelian-projected and center-projected variables, respectively. This allowed for the determination of the fraction of times $n$ faces of a cube dual to a magnetic-current link were pierced by $Z(3)$ flux. At $\beta=6$, it was observed that over 80$\%$ of the cases correspond to $n \geq 2$, with a predominance ($74\%$) of center-vortex pairs attached to monopoles ($n=2$). In  Ref. \cite{generalizedmag}, these authors noted that due to the inherent imprecision in methods for locating monopoles and P-vortices, the actual occurrence of $n\geq 2$ could potentially be as high as 100$\%$. For general $N$, it is interesting to note that nonoriented center-vortices ($n=2$) were recently shown to be a preferred saddle-point for YM theory in compactified spacetimes \cite{yuya}. 

The modeling of the mixed ensemble in Ref. \cite{mixed} was motivated
by the lattice description of an Abelian ensemble of worldsurfaces coupled to an external Kalb-Ramond field \cite{Rey}, where the Goldstone modes for the loop condensate are represented by an Abelian gauge link-variable $U_\mu \in U(1)$. In the center-vortex framework,  the Kalb-Ramond field can be used to express the intersection number between a center-vortex worldsheet $\mathcal{S}$ and a surface whose border is the Wilson loop $\mathcal{C}_{\rm e}$. This corresponds to the linking number between $\mathcal{S}$ and  $\mathcal{C}_{\rm e}$. In this manner, the average of center elements generated in Wilson loop averages over a center-vortex condensate can be represented by
gauge-field Goldstone modes with frustration. Building on this simple description, a generalized lattice gauge model with frustration was then proposed to generate the center element average. In this model, gauge link-variables $U_\mu \in SU(N)$ implement the natural matching property of center vortices in groups of $N$ and allow the inclusion of the nonoriented component through an ensemble of adjoint holonomies. Using polymer techniques, the latter gave rise to a set of effective adjoint Higgs fields. Of course, this is a phenomenological description where the starting point is not the pure YM theory for the chromoelectric link-variables $U^{(\rm e)}_\mu \in SU(N)$, but rather possible ensembles of topological objects detected in the YM Monte Carlo configurations. Similarly to replacing full Monte Carlo by projected averages, insights into the confinement mechanism can be gained by modeling averages over the objects identified in these projections. The characterization of such ensembles is also based on natural possibilities that could be explored in future lattice studies. Furthermore, their validation as potential mechanisms can be achieved by deriving their physical consequences and comparing them with additional lattice data to assess their feasibility. Along these lines, oriented and nonoriented center vortices were modeled within an Abelian-projected framework, using a wavefunctional peaked at the corresponding gauge fields \cite{wavefunctional}. By performing a functional Fourier transform, we switched to the dual electric field representation, where polymer techniques can be applied to integrate over the ensemble and represent the wavefunctional in terms of effective fields.
Being phenomenological in origin, the ensemble was constructed using elementary center vortices, which carry the Cartan fluxes $\beta_i$. In the monopole sector, we also considered the simplest elementary charges $\beta_i -\beta_j$, i.e., the magnetic weights of the adjoint representation. This minimal content allowed us to parametrize the main correlations while keeping the number of phenomenological parameters as small as possible, due to the physical equivalence between the different $\beta_i$. Although all the mass scales in this model are expected to be originated from a single nonperturbative mass scale in YM theory, the precise probabilities and weights for the different ensemble components would require different phenomenological parameters, which would increase with the inclusion of other Cartan charges. 
Recently, center vortices only characterized by their center charge $1, \dots , N-1$ were addressed in Ref. \cite{unsal}. Interestingly, the proliferation of monopoles sitting at the junction of $N$ charge $1$ center vortices in theories with $Z_N$ 1-form symmetry were shown to be crucial for confinement. One objective of this work is to derive an effective lattice representation for the partition function associated with an Abelian-projected ensemble. This will complement the effective representation previously obtained at the level of the wavefunctional. Again, we will keep a minimal content for the Cartan flux, based on the defining and adjoint weights of
$\mathfrak{su}(N)$.
Since the center-vortex Cartan fluxes involve $N$ equivalent possibilities $\beta_i$, we anticipate that this sector can be expressed in terms of a set of link-variables, with a weight symmetric under their permutation. A key question we will address in this context is how Abelian projection can be compatible with $N$-ality while simultaneously allowing for a ``dual superconductor'' description of the fundamental string. 
Another objective is to compare the Abelian-projected and non-Abelian mixed ensembles of center vortices and chains.
 To achieve this, we will employ a method capable of handling both scenarios: the Weingarten matrix representation of surfaces \cite{weingarten}. In particular, in both scenarios, this approach will help elucidate the emergence of gauge-field variables  as Goldstone modes in a phase where the center-vortex worldsurfaces percolate.

In Sec. \ref{mens}, we summarize the central physical inputs and previous developments regarding the modeling of mixed ensembles in four dimensions. In Sec. \ref{wsur}, we briefly review the noninteracting $N\times N$ matrix models
and interpret them as generating surfaces with $N$ possible labels at the vertices.  These models are stabilized by means of contact interactions in Sec. \ref{stabilized}, which also helps to clarify the emergence of gauge-field Goldstone modes when surfaces percolate.  In Sec. \ref{eloops}, lattice ensembles of loops and lines are briefly reviewed. The 4D lattice description and implications of the Abelian-projected ensemble, formed by  $N$ surface types with global magnetic charges, is given in Sec. \ref{ape}. In Sec. \ref{nae}, this is compared with the non-Abelian setting, where the magnetic charges are locally distributed at surface vertices.  Finally, Sec.  \ref{conclusions} is devoted to the conclusions.

\section{Mixed ensembles in four dimensions}
\label{mens}

Here, we  briefly summarize the main physical inputs about mixed ensembles of center vortices and chains. Some of them have already been detected in the lattice, while others are natural possibilities. Our approach is phenomenological, incorporating the collimated fluxes observed in lattice simulations.  When a collimated flux links the curve $\mathcal{C}_{\rm e}$, the Wilson loop $W_{\mathcal{C}_{\rm e}}$ gives a center element. Therefore, at asymptotic distances, when $\mathcal{C}_{\rm e}$ is large compared with the center-vortex thickness, we will model the Wilson loop average as a center-element average. To propose the simplest  ensemble components, it is instructive to examine the corresponding gauge field variables in the continuum, which can be parametrized by singular mappings $S \in SU(N)$ combined with suitable profiles to produce thick, collimated objects. This way, the natural correlations among these components can also be determined. The phase $S$ changes by a center element when encircling the vortex guiding-centers and characterizes the possible charges they carry (see Ref. \cite{universe} and references therein). Outside the center-vortex  localization region, where the field strength goes to zero, the gauge field $A_\mu^{\rm thick}$ can be approximated by:   
\begin{align}
    {\rm Ad}(a_\mu)=i {\rm Ad}(S) \partial_\mu {\rm Ad}(S^{-1})\;,
    \label{Adj}
\end{align}
which is a local (but not global) pure gauge. Here,
 ${\rm Ad}(\cdot)$ stands for the adjoint representation of $SU(N)$. This is a convenient parametrization as ${\rm Ad}(S)$ is single-valued when encircling the guiding-center and no  $\delta$-distributions originated from the multivaluedness of $S$ are manifested in $a_\mu$. Before listing some important examples, we note that a renormalizable procedure providing a first-principles foundation for the Yang-Mills (YM) ensemble in the continuum was developed in Refs. \cite{gabriel-oxman,singer1,singer2}.  Singer's no-go theorem \cite{singer-0} asserts the nonexistence of a global gauge fixing, thereby precluding the standard Faddeev-Popov (FP) quantization. However, the theorem does not rule out partitioning the configuration space $\{A_\mu\}$ and expressing the YM partition function as a sum over sectors where a standard FP treatment can be implemented. In Refs. \cite{gabriel-oxman,singer1,singer2}, a method similar to the lattice Laplacian center gauge \cite{dlcg} was used to propose a partition labeled by classes of singular mappings $S$, thus leading to a YM ensemble.  Nevertheless, an approximation scheme to derive the main properties of these sectors or to assess their potential representation in terms of collimated objects has not yet been developed. Addressing these important questions constitutes a challenging problem beyond the scope of this paper.

\subsubsection{Percolating center vortices}
\label{percolate}

The confining phase is formed by large clusters of percolating center vortices  \cite{percolating-1,percolating-2}.  In Refs. \cite{randomsu2, randomsu3}, a lattice random surface model that captures the order of the phase transition in $SU(2)$ and $SU(3)$ was proposed. This model is based on an action for oriented center-vortex worldsheets with stiffness, 
a property also observed in lattice simulations \cite{stiff1,stiff2}.  At the level of the partition function, as well as in the wavefunctional formalism, we have been modeling ensembles formed by elementary center vortices (center charge $\pm 1$). In the continuum, these configurations can be described by the mapping
\begin{align}
    S=e^{i\chi \, \beta\cdot T} \makebox[.5in]{,} \beta\cdot T\equiv \beta|_q T_q \;,
    \label{Scv}
\end{align}
where $\chi$ is multivalued and changes by $\pm 2\pi$ when going around the worldsurface $\mathcal{S}$ (guiding-centers),  $T_q$ are the diagonal (Cartan) generators of  $\mathfrak{su}(N)$, and $\beta$ is one of the $N$ magnetic weights $\beta_i = 2N \omega_i$ of the defining representation of $SU(N)$. The weights $\omega_i$, $i=1, \dots , N$, are formed by the elements of $T_q$ at the $i$-th position of the diagonal.\footnote{The  ``magnetic'' weights of a representation $D(\cdot)$ of $\mathfrak{su}(N)$ are given by $\beta_i =2N \omega $, where the weight $\omega$ is an $(N-1)$-tuple formed by eigenvalues of $D(T_q)$ for a simultaneous eigenvector $|\omega\rangle$:  $D(T_q)|\omega \rangle = \omega|_q |\omega \rangle$, $q=1, \dots, N-1$.}  In this case, Eq. \eqref{Adj} yields
\begin{align}
a_\mu = \beta\cdot T \, \partial_\mu \chi \;. \label{simple-vortex}
\end{align}
When a thick center vortex  is completely enclosed by a loop $\mathcal{C}_{\rm e}$, the group holonomy can be computed using $a_\mu$ instead of $A_\mu^{\rm thick}$. For a general group representation ${\rm D}(\cdot)$,
we have
  \begin{gather}
 	   {\rm D} \left(  e^{i \int_{\mathcal{C}} dx_\mu\, a_\mu(\mathcal{S})  } \right) = {\rm D} \left(  e^{i \Delta \chi  \,  \beta_i \cdot T   } \right) \;.
       \end{gather}
In addition, for  every $\beta_i$, we can write
       \begin{gather}
{\rm D} \left(  e^{i 2\pi  \,  \beta_i \cdot T   } \right)        = 	{\rm D} \left(  e^{-i \frac{2\pi }
 		{N} } I \right) =  e^{-i \frac{2\pi k}
 		{N} } I_{\mathscr D}\;,
 	\label{DPa2}
 \end{gather} 
which provides the same center-element for every $\beta_i$. Here, $k$ determines how the center of $SU(N)$ is realized in ${\rm D}(\cdot)$ ($N$-ality). The $N$-ality of the defining and adjoint representations are $k=1$ and $k=0$, respectively.  
 Then, when the flux is completely enclosed by $\mathcal{C}_{\rm e}$,
 the Wilson loop for heavy quark probes in an irreducible  representation ${\rm D}(\cdot)$, with dimensionality $\mathscr{D}$,  is given by
\begin{gather}  
W_{\mathcal{C}_{\rm e}}[A^{\rm thick}]   =  \frac{1}{\mathscr{D}} \, {\rm Tr}\, {\rm D}  \left(  e^{i \int_{\mathcal{C}_{\rm e}} dx_\mu\,  a_{\mu}(x)  }  \right) =  e^{-i \frac{2\pi k}
{N} L(\mathcal{S},\mathcal{C}_{\rm e})} I_{\mathscr D} \;,
\label{def-vortex}
\end{gather} 
where $ L(\mathcal{S},\mathcal{C}_{\rm e})$ is the 
linking number between the closed worldsurface $\mathcal{S}$ and the loop $\mathcal{C}_{\rm e}$.

\subsubsection{Center-vortex  $N$-matching}
\label{N-matching}

 As the defining magnetic weights 
satisfy
\[
\beta_1 + \dots + \beta_N = 0 \;,
\]
configurations characterized at a fixed time by $N$ lines $\gamma_1, \dots, \gamma_N$ carrying weights $\beta_1,\dots,\beta_N$ can be matched at a common endpoint
\cite{3densemble,universe,wavefunctional}.  For example, an array with two common endpoints can be described  in terms of the singular transformation
\begin{align}
    S=e^{i\chi_1 \beta_1\cdot T}\dots e^{i\chi_{N-1}\beta_{N-1}\cdot T}\;,
    \label{match}
\end{align}
where $\chi_i$ is an angle that is multivalued with respect to the worldsurface  spanned in time by the loops $\gamma_i\circ \gamma_N^{-1}$.
For $SU(3)$, the matching of three vortex lines carrying center charge $+1$ was observed at fixed-time slices in center-projected data \cite{branching,leinweber-nmatching}.

\subsubsection{  Center vortices  attached to  monopoles (collimated chains)}
\label{cha}

Another important class of configurations is formed by
collimated center-vortex fluxes attached in pairs to monopoles.
These objects, known as chains or nonoriented  center vortices (in the Lie algebra)
were detected in $SU(2)$ and $SU(3)$, with a procedure based on the Maximal Abelian gauge (MAG) \cite{Stack2002,collimation}.  Because of the monopoles, chains are essentially non-Abelian configurations \cite{conf-qg}. 
For example, the guiding-centers for a chain localized (at all times) on the $z$-axis with a monopole at $z=0$ is described by
\begin{align}
    S=e^{i\varphi \beta\cdot T}W(\theta)\makebox[.5in]{,} W(\theta)= e^{i\theta \sqrt{N}T_\alpha}\;,
    \label{Schain}
\end{align}
where $T_\alpha$ is an off-diagonal generator. 
This phase characterizes the transition from a center-vortex weight $\beta'$ to $\beta$, as we go from negative to positive $z$. This is mediated by a monopole with charge $\beta -\beta'=2N\alpha$. Indeed, at $\theta \sim 0$, $S \sim e^{i\varphi \beta\cdot T} $, while at  $\theta \sim \pi$, $S \sim W(\pi) \, e^{i\varphi \beta'\cdot T} $.

\subsubsection{Monopole worldline matching-rules}
\label{m-match}

In four dimensions, monopole worldlines carrying different adjoint weights can also be matched. For example, when $N \geq 3$, there is a phase $S \in SU(N)$ \cite{mixed} that represents three open worldsurfaces carrying magnetic weights $\beta$, $\beta'$, and $\beta''$, attached in pairs to open monopole worldlines carrying charges $\alpha$, $\gamma$, and $\delta$:
\begin{gather}
\beta- \beta' = 2N \alpha \makebox[.5in]{,} \beta' -\beta''= 2N \gamma \makebox[.5in]{,}  \beta'' - \beta = 2N \delta  \;.    
\end{gather}
In this configuration,  the three worldlines meet at their initial and final endpoints, thus forming a closed object. This matching is possible because of the condition
\begin{gather}
    \alpha + \gamma + \delta = 0 \;. 
    \label{m3}
\end{gather}
Of course, there are also higher order configurations involving, say, four monopole worldlines with charges that satisfy 
\begin{gather}
 \alpha + \gamma + \delta + \sigma = 0 \;. 
 \label{m4}
\end{gather}
These possibilities are the lower-dimensional counterpart of the matching of center-vortex worldsurfaces in groups of $N$, discussed in Sec. \ref{N-matching}.

\subsubsection{Non Abelian d.o.f.}

An interesting property that could be incorporated into the phenomenological approach is the presence of non-Abelian degrees of freedom (d.o.f.). Specifically, for each label $S$ in Eqs. \eqref{Scv}-\eqref{Schain}, while $S$ and $US$ describe the same sector, $S$ and $S \tilde{U}^{-1}$ represent physically inequivalent sectors, where $U$ and $\tilde{U}$ are regular mappings.  As these inequivalent defects are located at the same points in physical space, a continuum of non-Abelian d.o.f.  is implied.  These were modeled in center-vortex ensembles \cite{3densemble,mixed}, where they alter both the field content and the symmetries of the effective descriptions.

\vspace{.2cm}

In Secs. \ref{wavefun} and \ref{zf}, we shall discuss how the main ingredients listed above were implemented at the level of the wavefunctional and the partition function, respectively.  In both cases,  the objective was to obtain effective representations that could
 capture the confining properties of $SU(N)$ YM theory in asymptotically large Wilson loops, where the center-vortex thickness is only a boundary effect and $N$-ality properties should be reproduced. 
 
\subsection{The Abelian-projected mixed ensemble (wavefunctional)}
\label{wavefun}

One objective of this work is to represent, at the level of the partition function, an Abelian-projected ensemble of collimated objects carrying Cartan fluxes. This type of ensemble was previously explored in Ref. \cite{wavefunctional}, but through the derivation of the corresponding vacuum wavefunctional. A possible line of research would be to establish a direct relationship between them. As is well known, the vacuum wavefunctional can be obtained from a 4D partition function, path-integrated over configurations defined at $t \leq 0$, with an arbitrary fixed profile at $t=0$. This aspect will be pursued elsewhere. Instead, in Sec. \ref{ape}, we will model collimated fluxes around worldsurfaces and their correlations, such that, on a fixed time slice, they reproduce the configurations previously described in the wavefunctional formalism. In this section, within the Hamiltonian formalism, we will review the main steps leading to the vacuum wavefunctional $\Psi(A)$, representing the probability amplitude for the gauge field variable $A_i(\mathbf{x})$, $\mathbf{x} \in \mathbb{R}^3$, defined on a given time slice. 
To this end, as our primary focus was on computing Wilson loop averages in the asymptotic limit, we simply considered thin objects.  Thus, the ensemble of oriented center vortices was described by a 
wavefunctional peaked at gauge fields $a(\{\gamma\})$,
 \begin{align}
 \Psi(A)= \sum_{\{\gamma\}}\psi_{ \{\gamma \} }\, \delta (A-a(\{\gamma\}))   \label{sp2}\;.
\end{align}
The amplitude $\psi_{\{\gamma\}}$ 
parametrizes the center-vortex tension, stiffness, and interactions. The gauge field $a(\{\gamma\})$ is in the Cartan subalgebra of $\mathfrak{su}(N)$ and $\{ \gamma\} $ is a particular vortex network $\{\gamma\}$ formed by loops and a given distribution of defining magnetic weights $\beta_i$. It is obtained as a superposition of the gauge fields
\begin{align}
 a_\beta(\gamma) = 2\pi\beta \cdot T \,(-\nabla^2)^{-1}\nabla \times j(\gamma) \makebox[.5in]{,} j(\gamma) = \int_\gamma d\bar{x}\, \delta(x - \bar{x}) \;.\label{a3d}
\end{align} 
 For loops, $a_\beta(\gamma)$ represents a thin Cartan flux $\beta$ carried by $\gamma$, parametrized by $\bar{x}$. Note that this object is physical,
 as the corresponding Wilson loop is nontrivial and given by
 Eq. \eqref{def-vortex}, with the difference that in a fixed-time three-dimensional slice, the linking number is between $\mathcal{C}_{\rm e}$ and $\gamma$. On this slice, it also coincides with Eq. \eqref{simple-vortex}, with $\chi$ multivalued when going around $\gamma$.  
 The $N$-matching example in Sec. \ref{N-matching} corresponds to 
 \begin{gather}
 a(\{\gamma\}) = a_{\beta_1}(\gamma_1) + \dots + a_{\beta_N}(\gamma_N)  \;,
\label{Nex}
   \end{gather}
  where $\gamma_1, \dots, \gamma_N$ are open lines sharing common endpoints. In addition, the nonoriented thin vortex associated to the phase $S$ in Sec. \ref{cha} can be written as
\begin{align}
S_D \, a(\{\gamma\}) S_D^{-1}+iS_{\rm D}\nabla S_{\rm D}^{-1}\makebox[.5in]{,} a(\{\gamma\}) =a_{\beta}(\gamma)+a_{\beta'}(\gamma')+a_{\mathscr E}(\delta)) \makebox[.5in]{,}{\mathscr E} =\beta-\beta' \;,
\label{epsilon}
\end{align}
where the non-Abelian phase $S_{\rm D}$ is singular but single-valued when going around any loop.  Therefore, the Abelianized form of a nonoriented vortex corresponds to a Cartan flux $\beta$ ($\beta'$) carried by $\gamma$ ($\gamma'$) on the $z$-axis, starting (ending) at $z=0$. The flux is conserved because of an unobservable Dirac line $\delta$ ending at $z=0$. Indeed, Dirac lines are also characteristic of the De Grand-Touissant technique \cite{touissant} used to detect the monopole sector in the above mentioned lattice studies in the MAG. Unlike the Dirac string $\delta$, the center-vortex locations $\gamma$, $\gamma'$ cannot be altered by single-valued gauge transformations. Moreover, the Wilson loop computed with Eq. \eqref{epsilon} or its Abelianized form $a_\beta(\gamma)+a_{\beta'}(\gamma')+a_\mathscr{E}(\delta)$ give the same nontrivial result. In fact,  $a_\mathscr{E}(\delta)$ does not affect the Wilson loop, which only depends on the linking number between $\mathcal{C}_{\rm e}$ and the chain formed by $\gamma$ and $\gamma'$. To get rid of unphysical Dirac strings, only keeping the physical monopoles at their endpoints, the wavefunctional was cast in terms of a monopole potential $\zeta$ 
\begin{equation}
   \Psi (A,\zeta)=\sum_{\{\gamma\}}\psi_{\{\gamma\}}\, \delta\big(A-a(\{\gamma\})\big)\, \delta\big(\zeta-(-\Delta)^{-1}\nabla \cdot b(\{\gamma\})\big) \;,
   \label{b4}
\end{equation}
where $b(\{\gamma\})$ is a superposition 
of thin fluxes $b_\beta(\gamma)=2\pi\beta \cdot T j(\gamma)$. This potential allows us to define the variable $ \nabla \times A - \nabla \zeta$, which represents the chromomagnetic field. The effective description was attained in the dual language, switching from $\Psi (A, \zeta)$ to the electric-field representation   $\tilde{\Psi} (E, \eta)$ via a functional Fourier transform.  In this manner, we obtained
\begin{align}
    \tilde{\Psi}(E,\eta)=\sum_{\{\gamma\}}\psi_{\{\gamma\}}e^{i\sum\int_{\gamma}dx\cdot\Lambda_\beta}\makebox[.5in]{,}\Lambda_\beta  = 2\pi \beta \cdot \Lambda \makebox[.5in]{,} \Lambda = \frac{\nabla\times E}{-\nabla^2}+\frac{\nabla\eta}{-\nabla^2}\;. \label{starting-pt}
\end{align}
The sum in the exponent refers to a summation over the line components $\gamma$ and the corresponding distribution of weights among the $\beta_i$ within a given network $\{\gamma\}$. 
Next, we used a phenomenological probability amplitude $\psi_{\{\gamma\}}$ with phenomenological  parameters $\mu$ (tension) and $\kappa^{-1}$ (stiffness), in accordance with lattice simulations \cite{stiff1,stiff2}. In this expression, the main building-block is the sum over all the lines carrying a given $\beta$, with fixed initial/final points and orientations, as well as fixed length $L$:
\begin{align}
 \sum_\gamma e^{-\int_0^L ds\,\left(\frac{1}{2\kappa}\dot{u}(s) \dot{u} (s)+\mu-i\sigma(x) -i\dot{u}(s)\cdot \Lambda_{\beta}(x(s))\right)}\makebox[.5in]{,} \dot{u}(s) \equiv \frac{d x(s)}{ds}\;.
\end{align}
The   auxiliary scalar field $\sigma(x)$ was introduced to encode contact interactions between the vortex lines, implemented through integration with a Gaussian weight. This building-block satisfies a Fokker-Planck diffusion equation, whose solution in the limit of small stiffness is  
\begin{align}
    \langle x| e^{- L O_{\beta}} |x_0 \rangle\makebox[.5in]{,} O_{\beta} =-\frac{1}{3\kappa}D^2(\Lambda_\beta)+\mu-i\sigma(x)\makebox[.5in]{,}D(\Lambda_\beta)= \nabla - i\Lambda_\beta\;.
\end{align}
Then, for fixed endpoints, the integral over the lengths $L$ gives a two-point function
\begin{align}
 O^{-1}_{\beta}(x,x_0)\;,
\end{align}
and the sum over vortex networks in Eq. \eqref{starting-pt} becomes an effective field theory. Namely, the center-vortex loops generate a functional determinant, which can be exponentiated by introducing $N$ complex scalar fields $\phi_1, \dots , \phi_N$, corresponding to the possible fluxes $\beta_1,\dots, \beta_N$. In addition, the different connections between open lines correspond to Feynman vertices. 
 Organizing the fields as an $N\times N$ diagonal matrix $\Phi$ whose entries are the complex scalars $\phi_i$, we obtained  
 \begin{align}
     &\tilde{\Psi}(E,\eta)=\int D\Phi
     \, e^{-W[\Phi,\Phi^\dagger,\Lambda]}\makebox[.5in]{,} 
     W[\Phi,\Phi^\dagger,\Lambda]=\int d^3x \left({\rm Tr} ((D(\Lambda)\Phi^\dagger)D(\Lambda)\Phi)+V(\Phi,\Phi^\dagger)\right)\;,\nonumber\\&
V(\Phi,\Phi^\dagger)= \frac{\lambda}{2}{\rm Tr}(\Phi^\dagger\Phi-\vartheta^2 I_N)^2-\xi(\det\Phi+\det\Phi^\dagger)- \nu  {\rm Tr}(\Phi^\dagger T_A \Phi T_A)\;.
\label{wavef}
 \end{align} 
The parameter $\lambda$ gives the strength of the repulsive contact interaction among center vortices. In addition, 
$\xi$ and $\nu$ weight the importance of $N$-matching and the monopole component in chains, respectively. Note that, in the last term, the partial sum over the Cartan sector $A=q$ only redefines the mass parameter, while the off-diagonal contribution is proportional to $\sum_{i\neq j} \bar{\phi}_i\phi_j $. This precisely generates the transitions between center-vortex lines labeled by $\beta_i$ and $\beta_j$, thus introducing  chains into the ensemble. The first 
term is $U(1)^{N}$-symmetric. Then, if percolation were the only effect 
($\vartheta^2 > 0$), no domain wall could be formed as the vacuum manifold would be $U(1)^N$, which is not discrete.  When percolation dominates and $N$-matching is strong, because of the $\det \Phi$-term, the most relevant degrees are on $U(1)^{N-1}$.  Finally, turning-on the monopoles to form collimated chains ($\nu \neq 0$), the symmetry is reduced to $Z(N)$ and the vacua become discrete.
This resulted in confinement through the formation of a domain-wall localized on the Wilson loop \cite{wavefunctional}, as revealed when computing the average
\begin{align}
  \langle W_{\mathcal{C}_{\rm e}}\rangle = 
  \int [DA][D\zeta]\, |\Psi(A, \zeta)|^2 \, W_{\mathcal{C}_{\rm e}}[A,\zeta] 
\end{align}
in the dual language. 
 The asymptotic string tension thus obtained satisfies the Casimir law.

\subsection{Mixed ensemble with non-Abelian d.o.f. (partition function)}
\label{zf}

 In Ref. \cite{mixed} (see also \cite{universe}), the construction of the 4D partition function for center vortices was motivated by the lattice description of the Abelian ensemble of surfaces given in Ref. \cite{Rey}. In the latter, a complex-valued field $V(\gamma)$ defined on closed loops $\gamma$ was  considered. In particular, it was argued that the loop condensate, where $V(\gamma)$ has frozen modulus, is described by the  Abelian Wilson action for link variables $U_\mu(x) \in U(1)$. In other words, the Goldstone modes for a condensate of Abelian loops are given by a $U(1)$ gauge-field. Moreover, the Kalb-Ramond field $b_{\mu \nu}$ coupled to the surfaces  manifests as a frustration at the plaquettes given by $e^{i a^2 b_{\mu\nu}}$. Now, in the center-vortex framework, when the linking number between a surface and a Wilson loop is written as an intersection number, the associated Kalb-Ramond field becomes concentrated on an arbitrary surface $\mathcal{S}_{\rm e}$ whose border is the Wilson loop $\mathcal{C}_{\rm e}$. 
In the lattice, when considering elementary center vortices, this corresponds to a frustration $e^{i b(p)}$, where $b(p)= \pm 2\pi k/N$ if $p$ intersects $\mathcal{S}_{\rm e}$
and it is trivial otherwise. The sign depends on the relative surface orientations and $k$ is the $N$-ality of the quark representation ${\rm D}(\cdot )$.  In this manner, we proposed the representation
\begin{gather}
\frac{Z_{\rm v}^{\rm latt} [b]}{Z_{\rm v}^{\rm latt} [0]}  \makebox[.5in]{,} Z_{\rm v}^{\rm latt} [b] = 
 \int [{\cal D} U_\mu] \, e^{-S^{\rm latt}_{\rm v}(b) } \;,  \nonumber \\
 S^{\rm latt}_{\rm v} (b) \propto  \sum_{\mathbf{x}, \mu < \nu } \mathrm{Re}\;  {\rm Tr} \left[ I - e^{i b(p)} U_\mu(x) U_\nu(x + \hat{\mu}) U^{-1}_\mu(x + \hat{\nu}) U^{-1}_\nu(x)   \right] \;, 
 \label{Wf}
\end{gather} 
 to generate a center-element average when a Wilson loop in representation  ${\rm D} $ is in the presence of an ensemble of percolating and oriented center-vortex worldsurfaces. In Ref. \cite{mixed}, driven by the possible presence of non-Abelian degrees of freedom (d.o.f.) on center vortices, a starting point based on non-Abelian variables $U_\mu(x) \in SU(N)$ was considered. This way, surfaces can also meet in groups of $N$ at common links forming a closed path ($N$-matching). In addition, chains were included by means of a ``gas'' of dual adjoint holonomies. Weighting them with the simplest geometrical properties (tension and stiffness), we obtained the lattice mixed ensemble of center vortices and chains 
 \begin{gather}
 Z^{\rm latt}_{\rm mix} [b] =   \int [{\cal D} U_\mu] \, e^{- 
 S^{\rm latt}_{\rm v} (b)   } \, \times Z_{\rm m}  \;,
\label{mixZ}
\end{gather}
where $Z_{\rm m}$ is the partition function for the monopole component. The naive continuum limit gave place to an effective $SU(N)$ gauge field $\Lambda_\mu$ that represents the percolating center-vortex worldsurfaces. 
In addition, the monopole ensemble was integrated by using polymer techniques, which led to minimally coupled interacting adjoint scalar fields. The natural matching rules between the monopole worldlines correspond to Feynman vertices.  Finally, this, together with the emergent non-Abelian Goldstone modes $\Lambda_\mu$, led to a class of effective Yang--Mills--Higgs (YMH) models. Further details about this construction will be provided and clarified in the following sections, within the framework of the Weingarten representation.

\section{Matrix representation of surfaces}
\label{wsur}

 We start by reviewing 
 the lattice ensemble of  noninteracting closed oriented surfaces $s$ (connected or not)  proposed in  Ref. \cite{weingarten}\footnote{The sum over $\mathcal{S}$ includes all possible  orientations. }  
    \begin{align}
        Z_0 =\sum\limits_{ \mathcal{S}}  N^{\chi(\mathcal{S})} e^{-\sigma A(\mathcal{S}) }  \;.
        \label{gen-supens}
    \end{align}
    Here,  $A(\mathcal{S})$ and $\chi(\mathcal{S})$ are  the area and the Euler characteristic of $\mathcal{S}$, $N$ is natural, and $\sigma$ 
is a tension parameter.  This partition function was shown to be equivalent to a model where the variables are $N\times N$ dimensional matrices $V(x,y)$ of complex numbers that live at the links and satisfy $V(y,x)=V^\dagger(x,y)$. More precisely, there is a $\gamma$ such that
\begin{gather}
    Z_0=\int DV\, \exp\left(\gamma\sum_{p}{\rm Tr}V(p)-Q_0[V]\right)
   \;,\label{matrixmodel}
\\
V(p)= V(x,y)V(y,z)V(z,w)V(w,x) \makebox[.4in]{,} Q_0[V] =   \sum_{\{ x,y\}} \mathcal{Q}_0(V(x,y)) 
\makebox[.4in]{,} \mathcal{Q}_0(V) = {\rm Tr} \left(  V^\dagger V \right) \;,
\\
 DV =\prod\limits_{\{x,y\}} dV(x,y)
 \makebox[.5in]{,} dV(x,y) = 
 \prod\limits_{i,j}\pi^{-1}d[{\rm Re}\, V_{ij}(x,y)]d[{\rm Im}\, V_{ij}(x,y)]   \;,
 \label{wein}
\end{gather}  
where $x,y,z,w$ are the vertices of a plaquette.
The plaquettes include both orientations, that is, for every $V(p)$ there is also a variable $V(p') =V^\dagger(p)$. On the other hand, $\{ x,y\}$ denotes a nonoriented link. We will see that the partition function $Z_0$ can also be thought of as an ensemble of noninteracting colored surfaces, where $N$ colors are independently assigned to each vertex.   By expanding Eq. \eqref{matrixmodel} in powers of $\gamma$,   the contribution of  $F$ different plaquettes $p_1,  \dots, p_{F}$ is
\begin{equation}
 C(p_1, \dots, p_F) = \frac{\gamma^{F}}{F!}  \int DV\, ({\rm Tr}V(p_1)+ \dots + {\rm Tr}V(p_{F}))^{F}  e^{-Q_0[V]}\;.
  \label{contri}
\end{equation}
There are $F$ manners to choose a plaquette from the first factor in Eq. \eqref{contri}, $F-1$ from the second, etc., so that
\begin{align}
  &C(p_1, \dots, p_F) = \gamma^{F}  \int DV\,   {\rm Tr}V(p_1) \dots  {\rm Tr}V(p_{F})  \, e^{-Q_0[V]}\label{contri} \;.
\end{align}
 For the Gaussian measure,
\begin{align}
    \int DV\,  V_{o_1p_1}(x_1,y_1)\dots V_{o_{2n}p_{2n}}(x_{2n},y_{2n}) \, e^{-Q_0[V]}=\nonumber\\
    \sum_{(j_1,\dots,j_n),~ (k_1,\dots,k_n)}\delta_{o_{j_1}p_{k_1}}\delta_{x_{j_1}y_{k_1}}\delta_{y_{j_1}x_{k_1}}\dots\delta_{o_{j_n}p_{k_n}}\delta_{x_{j_n}y_{k_n}}\delta_{y_{j_n}x_{k_n}} \label{factorization}\;,
\end{align}
where $(j_1,\dots,j_n)$, $  (k_1,\dots,k_n)$ are permutations of the sequence ($1,\dots,n$).  
If there is no manner to group all the links in pairs having the same location but different orientations, then the result is zero.  The nontrivial contributions arise when the plaquettes can generate closed oriented surfaces
with $F$ faces.  Each pairing defines how to move on the surface to follow the orientation and how to assign the colors.  In other words, 
the partition function in Eq. \eqref{matrixmodel} generates the ensemble ($a$ is the lattice spacing)
\begin{align}
    Z_0 = \sum_{ \mathcal{S}_{\rm c}} e^{- \mu_0  A(\mathcal{S}_{\rm c})}   \makebox[.5in]{,} \gamma = e^{-\mu_0 a^2} \makebox[.5in]{,}
    A(\mathcal{S}) = a^2 F
    \;,
    \label{gene}
\end{align}
where $\mathcal{S}_{\rm c}$ denotes the different oriented colored surfaces. Regarding the distribution of colors, let us initially consider plaquettes $  p_1, \dots , p_F$ with single occupation at the common links. This means that there is only one variable of type $V(x,y)$ (and one of type $V(y,x)$). 
When $N=1$, $V(x,y)$ is a complex number and the integral 
\begin{align}
  \int dV(x,y)\, |V(x,y)|^2\, e^{-|V(x,y)|^2}  = 1 \;,
    \label{spair}  
\end{align} 
leads to just one surface. 
For general $N$ and single occupation, there is still only one orientation induced by the plaquettes that participate in its construction.\footnote{ There is also a contribution with the same geometry but different orientation obtained from another term in the expansion containing the adjoint variables.} On the other hand, as the contribution originates from products of the from  $V_{ij}(x,y) V_{kl}(y,x)$ when $j=k$ and $i=l$, $N$ possible labels are implied at each one of the $V$ vertices:
\begin{align}
 C(p_1, \dots, p_F) = N^V \gamma^F \;.
\end{align}

\subsection{Growing a surface}

\begin{figure}
    	\includegraphics[scale=1]{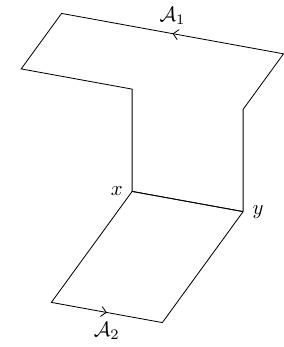}
    \caption{The surfaces $\mathcal{A}_1, \mathcal{A}_2$ with a common link $(x,y)$.}
    \label{fig3}
\end{figure}

\begin{figure}
\centering
\begin{subfigure}{.3\textwidth}
  \centering
  \includegraphics[scale=1.1]{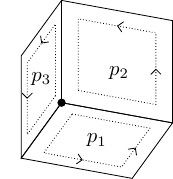}
  \subcaption{$n=3$}
\end{subfigure}%
\begin{subfigure}{.3\textwidth}
  \centering
    \includegraphics[scale=0.9]{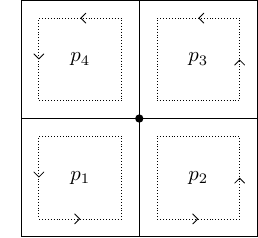}
     \subcaption{$n=4$}
\end{subfigure}
\begin{subfigure}{.3\textwidth}
  \centering
    \includegraphics[scale=1.0]{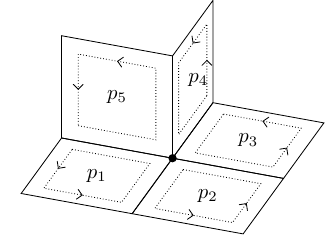}
     \subcaption{$n=5$}
\end{subfigure}
\caption{Oriented plaquettes meeting at a common vertex. }
\label{fig4}
\end{figure}

In order to discuss more general configurations, it is convenient to grow a surface by gluing simpler elements. For this aim, it will be useful to consider the path-ordered product (``holonomy'')
\begin{gather}
\Gamma(\mathcal{C}) = V(x_1 , x_2) V(x_2 , x_3) \dots  \;, 
\label{holog}
\end{gather}
 where the links $(x_1, x_2)$, $(x_2 , x_3)$, $\dots$ form the curve $\mathcal{C}$. 
When two holonomies along closed paths $\partial \mathscr{A}_1$, $ \partial \mathscr{A}_2$ have a common link $(x,y)$ with single occupation and traveled with opposite orientations (see Fig. \ref{fig3}), using 
\begin{align}
\int dV \,   {\rm Tr} (V A)   {\rm Tr} (V^\dagger B) \, e^{-\mathcal{Q}_0(V)} = {\rm Tr} (AB) \;,
\label{VXY} 
\end{align}
we have
\begin{equation}
\int dV(x,y) \,   {\rm Tr} (\Gamma(\partial \mathscr{A}_1))   {\rm Tr} (\Gamma(\partial \mathscr{A}_2)) \, e^{-\mathcal{Q}_0(V(x,y))}  = {\rm Tr} (\Gamma(\partial (\mathscr{A}_1 \cup \mathscr{A}_2)))   \;.
\label{VXY'}
\end{equation}
Then, consider an open surface element formed by $n$ plaquettes glued at links with single occupation that share just one common vertex (see Fig. \ref{fig4}).  
Using Eq. \eqref{VXY'} 
 to integrate one variable shared by $p_1 , p_2$, one shared by $p_2 ,p_3$, up to one shared by $p_{n-1},p_n$, we are left with an integral of the form: 
\begin{align}
    \int dV\, {\rm Tr}(V V^\dagger A) \, e^{-\mathcal{Q}_0(V)} = N {\rm Tr}(A) \;,
 \label{Nt}
\end{align}
computed on the link shared by $p_n ,p_1$. Thus, the integrals glue the plaquettes and give a partial contribution in terms of the holonomy along the border of the surface element, which contains the nonintegrated variables. This reproduces the factor $N$ for each vertex of $s$, discussed above:
\begin{align}
     C(p_1, \dots, p_n) = N {\rm Tr}(\Gamma (\partial \mathscr{A} )) \makebox[.5in]{,} \mathscr{A} = p_1 \cup  \dots \cup  p_n \;.
\label{GA}
\end{align}

\subsection{A closer look at surfaces in contact}
\label{tsur}

\begin{figure}[t]
	\centering
	\begin{subfigure}[b]{0.3\textwidth}
		\centering
		\includegraphics[scale=1]{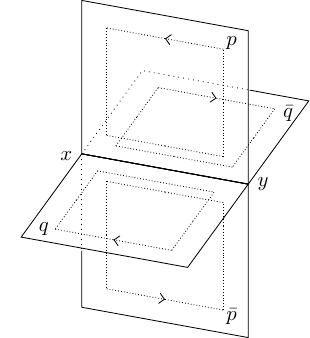}
	\subcaption{Array formed by $p,\bar{p},q,\bar{q}$.} 
 \label{fig7:a}\end{subfigure}
 \begin{subfigure}{.3\textwidth}
		\centering
		\includegraphics[scale=1]{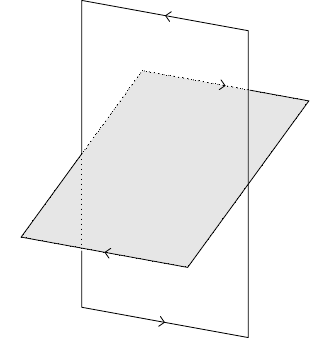}
	\subcaption{ $p$ ($q$) is glued with $\bar{p}$ ($\bar{q}$).}
 \label{fig7:b}
 \end{subfigure}%
	\begin{subfigure}{.3\textwidth}
		\centering
			\includegraphics{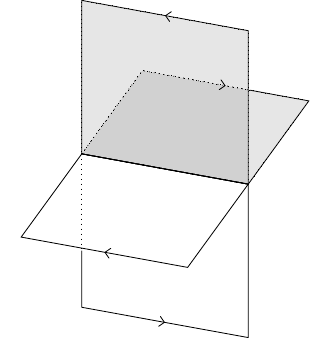}
	 \subcaption{$p$ ($q$) is glued with $\bar{q}$ ($\bar{p}$).}
\label{fig7:c}
\end{subfigure}
	\caption{The gluing of four plaquettes: noninteracting and general $N$  case }
 \label{fig7}
\end{figure}

Let us take a closer look at surfaces in contact, which occurs at links with higher occupation.  For example,  consider configurations generated from $F$ plaquettes $  p_1, \dots , p_F$, with $E$ common links, where $E_1$ of them have variables with single occupation and the double links are distributed along a curve $\mathcal{C}$ of total size $E_2$ ($E = E_1 + E_2$).   
For $N=1$,  the variables at doubly  occupied links appear in the form $V^2(x,y) V^2(y,x)$, which yields
\begin{align}
      \int dV\, (V \bar{V})^2 \, e^{-|V(x,y)|^2} =2   \makebox[.5in]{,} C(p_1, \dots, p_{F}) =  2^{E_2} e^{-\mu_0 A(\mathcal{S})}    \;. 
      \label{semb}
\end{align}
This corresponds to the $ 2^{E_2}$  different oriented surfaces that can be constructed out of  $p_1, \dots, p_F$. For general $N$, each pairing in Eq. \eqref{factorization}\ defines a choice to glue the oriented plaquettes or, how to move on the surface to follow the orientation. The important point is that the number of vertices and edges of the generated surfaces is generally different from the number of lattice sites and links that participate.  Consider an array of four plaquettes $p$, $\bar{p}$, $q$, $\bar{q}$ that meet at a common link $(x, y)$ with double occupation (see Fig. \ref{fig7:a}). In the $\gamma$-expansion (cf. Eq.  \eqref{contri}), the integration over the variables at $(x,y)$ gives place to two terms (cf. Eq. \eqref{a4}). Each one of them is the product of a pair of holonomies along the border of the surface elements $p \cup \bar{p}$, $q \cup \bar{q}$ and  $p \cup \bar{q}$, $q \cup \bar{p}$ depicted in Figs. \ref{fig7:b} and \ref{fig7:c}, respectively. 
In Appendix \ref{doc}, we consider consecutive arrays of this type along ${\mathcal C}$. For each one of the $2^{E_2}$ combinations of terms, we have to integrate over the common links to determine the generated surfaces. This allows to visualize whether a given site (link) on $\mathcal{C}$ contains a single vertex (edge) or a pair of superimposed surface vertices (edges). For a given orientation, the number of vertices $V(\mathcal{S})$, edges $E(\mathcal{S})$, and faces $F(\mathcal{S})$ of $\mathcal{S}$ turns out to be 
\begin{align}
    & V(\mathcal{S}) = V_1+ 2V_2 \makebox[.5in]{,} E(\mathcal{S}) = E_1+2E_2 = 2F  \makebox[.5in]{,} 
    F(\mathcal{S}) = F \;, 
\end{align}
where $V_1$ ($V_2$) is the number of sites that contribute with a factor $N$ ($N^2$). That is, there are $N$ possible colors for each vertex of $\mathcal{S}$. 
 A similar situation applies to higher-order occupations, so that
\begin{align}
    Z_0 = \sum_{ \mathcal{S}_{\rm c}} e^{- \mu_0  A(\mathcal{S}_{\rm c})} = \sum_{ \mathcal{S}} N^{V(\mathcal{S})} e^{- \mu_0  A(\mathcal{S})}
    \;. \label{sum-colors}
\end{align} 
 Note that because of the lattice constraint $  E(\mathcal{S}) = 2F $,  the degeneracy can be rewritten by redefining the tension and using the Euler characteristic  $\chi(\mathcal{S})$ of $\mathcal{S}$:
\begin{eqnarray}
	N^{V(\mathcal{S})}  \gamma^{F} =  N^{\chi(\mathcal{S})} e^{- \sigma A(\mathcal{S}) } \makebox[.5in]{,} \sigma = \mu_0  - a^{-2} \ln N \makebox[.5in]{,}    \chi(\mathcal{S}) =V(\mathcal{S})-E(\mathcal{S}) +F(\mathcal{S})   \;.  
\end{eqnarray}
This is the content encoded in Eq. \eqref{gen-supens}, which includes all surface types that intersect and/or touch along some curves. Among the orientations, there are disconnected cases where it is not possible to follow the gluing of the plaquettes so as to cover the whole surface.  Then, the partition function for noninteracting surfaces can also be cast in the form \cite{kawai1981}
\begin{align}
    Z_0 = e^{W_0} \makebox[.5in]{,}   W_0 = \sum_{\mathcal{S} \in \{ S_{\rm conn} \} } N^{\chi(\mathcal{S})} e^{-\sigma A(\mathcal{S}) } \;,
\end{align}
where the sum is over the set of connected oriented surfaces, for which the genus $g(\mathcal{S}) = (2-\chi(\mathcal{S}))/2 $ can also be used.

\section{The Goldstone modes for percolating surfaces}
\label{stabilized}

\begin{figure}[t]
\centering
\begin{subfigure}{.5\textwidth}
  \centering
\includegraphics[scale=0.5]{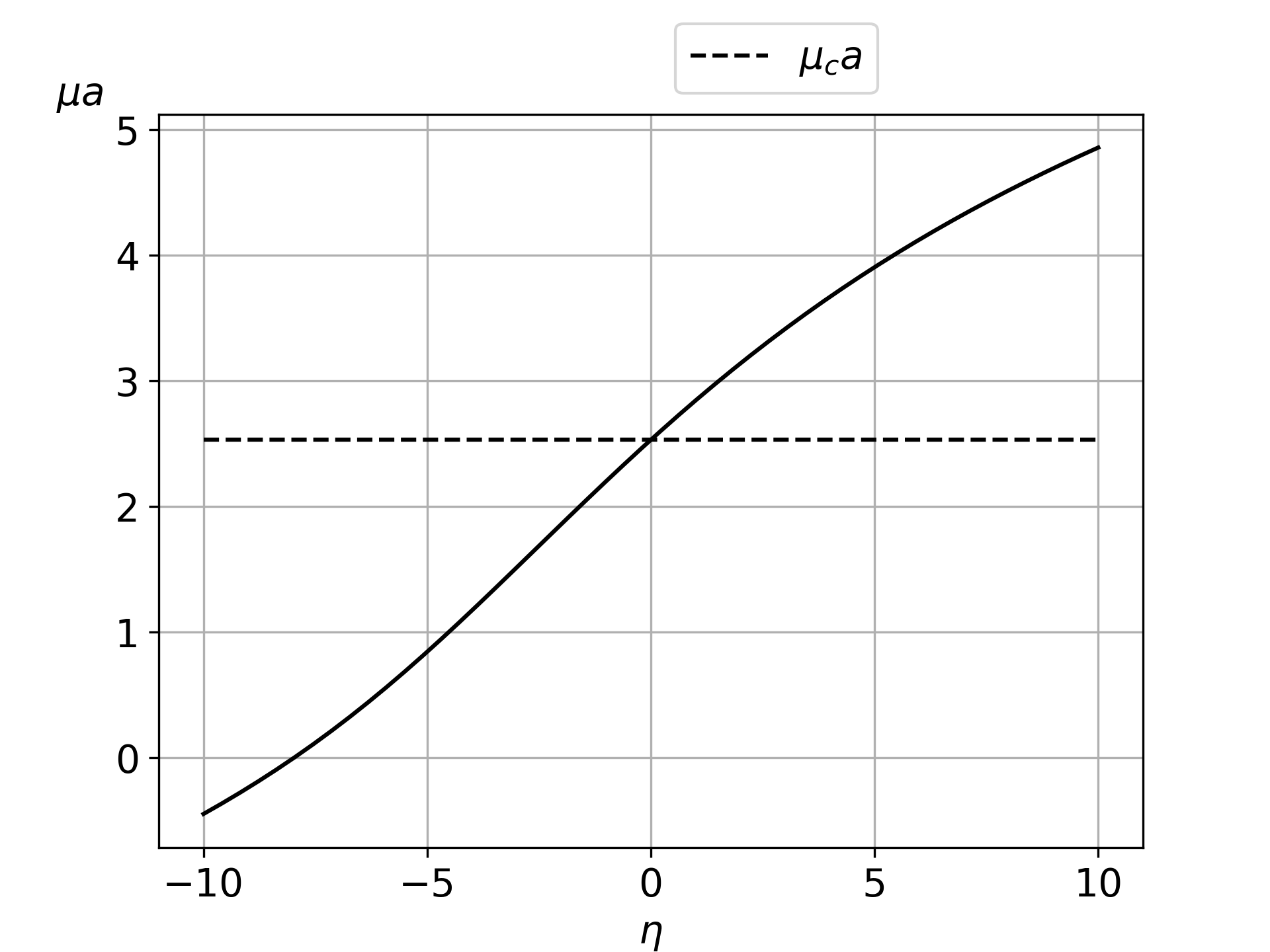}
\end{subfigure}%
\begin{subfigure}{.5\textwidth}
  \centering
\includegraphics[scale=0.5]{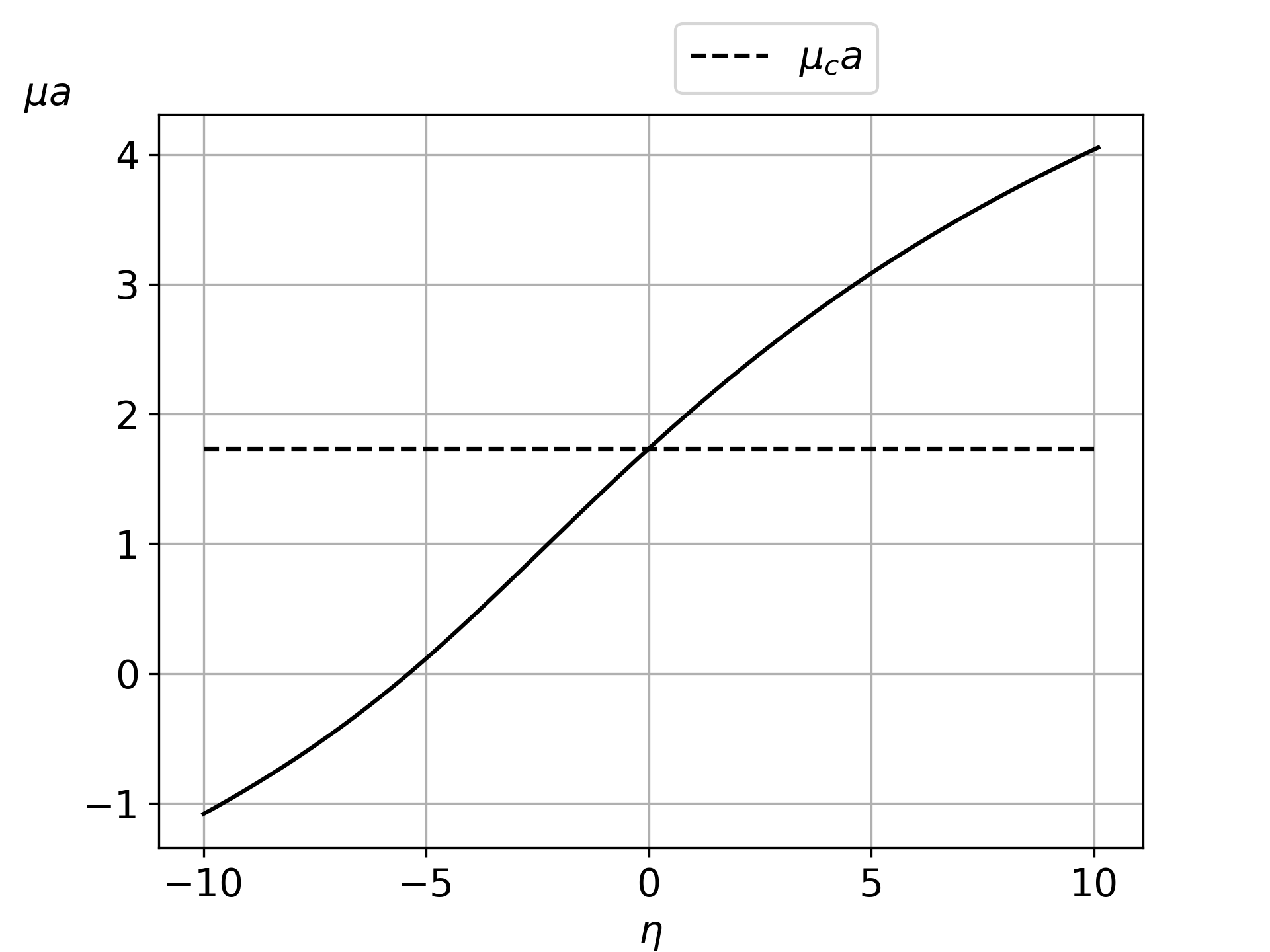}
\end{subfigure}
\caption{The renormalized tension $\mu$ as a function of $\eta$ for $\lambda=4, \gamma=1$.  When $\mu < \mu_{\rm c}$, a condensate is formed ($\eta<0$). This is shown for $N=1$ (left) and $N=3$ (right). }
\label{fig:condensate}
\end{figure}

As discussed in Ref. \cite{weingarten}, even in a finite lattice with periodic boundary conditions, the model defined by Eq. \eqref{matrixmodel}  is pathological, as the integral over the link variables gives a divergent $Z_0$. In that reference, sixth-order or higher-order interaction terms were proposed to stabilize the model. Its stabilization with a quartic term is also possible. Indeed,
the Von Neumann trace inequality,
\begin{equation}
    |{\rm Tr} (AB)| \leq \sum_{i=1}^N \sigma_i (A) \sigma_i (B) \;,
\end{equation}
where $\sigma_1 \geq \sigma_2 \geq \dots \geq \sigma_N$ are the singular values of an $N\times N $ complex matrix, leads to
\begin{equation}
    |{\rm Tr} (AB)|  \leq \frac{1}{2} \sum_{i=1}^N \left(\sigma_i^2 (A) + \sigma_i^2 (B) \right)=
    \frac{1}{2} {\rm Tr} (A^\dagger A + B^\dagger B)\;.
\end{equation}
For a generic plaquette $V(p)= ABCD$, this implies
\begin{eqnarray}
{\rm Tr} V(p) \leq |{\rm Tr} V(p)| \leq 
  \frac{1}{2} {\rm Tr} ((AB)^\dagger (AB) + (CD)^\dagger (CD))
  \nonumber \\
  =  \frac{1}{2} {\rm Tr} ( A^\dagger AB B^\dagger) + \frac{1}{2} {\rm Tr}  (C^\dagger CD D^\dagger) \nonumber \\
  \leq  \frac{1}{4} {\rm Tr} ( (A^\dagger A)^2 + (B^\dagger B)^2 + (C^\dagger C)^2 + (D^\dagger D)^2) \;.\label{thebound}
\end{eqnarray}
Note that starting from a given $p$ in  Eq. \eqref{thebound} and summing over the oriented plaquettes of the lattice that are in the same plane, each nonoriented link appears four times. As a given link can be in 3 different planes, we get the following bound
\begin{align}
    \sum_{p}{\rm Tr}V(p)  \leq 
3 \sum_{\{x,y\}}   {\rm Tr} (V^\dagger (x,y)V(x,y))^2  \;.
\label{ebound}
\end{align}
 Therefore, let us consider the quartic model ($n_l$ is the total number of links)
\begin{align}
    Z&= \frac{1}{\mathcal{N}(\lambda)}  \int DV\, \exp\left(\gamma\sum_{p}{\rm Tr}V(p)-Q[V]\right) \makebox[.5in]{,} \mathcal{N}(\lambda) = z^{n_l} \;,\label{quarticmodel} \\
 Q[V] =   \sum_{\{x,y\}} \mathcal{Q}(V(x,y)) &  \makebox[.4in]{,} \mathcal{Q}(V) = {\rm Tr}\left( \eta V^\dagger V+\lambda  (V^\dagger V)^2\right) \makebox[.4in]{,} z = \int dV\, e^{- \mathcal{Q}(V)}  \;.
\label{Qmodel}
\end{align} 
 The factor $\mathcal{N}(\lambda)$ guarantees that, in the $\gamma$-expansion, all the links that do not participate in a given surface lead to a trivial contribution. Because of Eq. \eqref{ebound}, the partition function 
 $Z$, where all the link variables are correlated, is bounded by an uncorrelated integral: 
\begin{align}
Z \leq  \frac{\mathcal{N}(\lambda')}{\mathcal{N}(\lambda)}  \makebox[.5in]{,} \lambda' = \lambda -3\gamma \;.
\end{align}
 Then, in the region  $\lambda > 3 \gamma$, the model is no longer pathological. For single occupation, the only change is a factor $\alpha$ in the right-hand side of Eqs. \eqref{VXY}-\eqref{Nt}
\begin{eqnarray}
& &  \alpha = \frac{1}{z N^2} \int dV \, {\rm Tr} ( VV^\dagger)\,  e^{-\mathcal{Q}(V)} \;.
\end{eqnarray} 
In this case, $F$ plaquettes leading to surfaces with $V$ vertices and $E$ edges yield
\begin{align}
 C(p_1, \dots, p_F) =  N^V \gamma^F \alpha^{E}   =   N^V  e^{-\mu A(\mathcal{S}) } 
 \makebox[.5in]{,}   e^{-\mu a^2 } =  \gamma \alpha^2\;,
\end{align}
that is, the tension is renormalized. 
Now, as is well known, the Weingarten model for surfaces does not correspond to a field theory \cite{kawai1981}. However, this applies to the normal phase.
As we shall see, this situation changes in the percolating phase. In effect, after adding and subtracting the bound 
\eqref{ebound} in the exponent of Eq. \eqref{quarticmodel},  we can also write the partition function in the form
\begin{align}
& Z= \frac{1}{\mathcal{N}}  \int DV\, \exp\left(-S[V]\right) \makebox[.5in]{,} S[V] = K[V] + U[V]\;, 
\end{align}
\begin{align}
& K[V] = 3 \gamma\sum_{\{x,y\}}  {\rm Tr} ( V^\dagger (x,y)V(x,y))^2  -\gamma\sum_{p}{\rm Tr}V(p) \;, \nonumber \\
& U[V]=\lambda'  \sum_{\{x,y\}} {\rm Tr} \,  (V^\dagger(x,y) V(x,y) -\vartheta^2 I)^2    + c \makebox[.5in]{,} 
    \vartheta^2 = \frac{-\eta}{2\lambda'} \makebox[.5in]{,} 
c = -\lambda' N \vartheta^4 n_l \;.
\end{align}
  By construction, $K[V] $ is positive definite. This also applies to the squared terms in $U[V]$. In Fig. \ref{fig:condensate}, we see that
when the renormalized tension is below a critical value  $\mu_{\rm c} >0$ (percolating surfaces), the parameter $\eta $ becomes negative, thus leading to a transition where $\vartheta^2$ becomes positive.  The absolute minima of $S[V]$ are those minima of $U[V]$ that nullify $K[V]$. The former condition means 
\begin{align}
    V(x,y) = \vartheta\,  U(x,y) \makebox[.5in]{,} U(x,y) \in U(N)  \;.
    \label{freeze}
\end{align}
At these minima, in order for $K[V]$ to be zero, the bound \eqref{thebound}    must be saturated for every plaquette. Then,  at the absolute minima, all the plaquette variables $U(p)$ formed out of the link variables $U(x,y)$ must be trivial. That is, $U(x,y)$
must be pure gauge.  Now, in the thermodynamic limit, a condensate dominated by the ``Goldstone'' modes in Eq. \eqref{freeze} is expected to be formed, accompanied by massive fluctuations away from the minima
of the potential, with ``mass'' $\lambda' \vartheta^2$. 
In the case where $\lambda'$ is much larger than $\gamma$, the massive modes get suppressed, we can freeze the ``modulus'' of $V(x,y)$, and keep the softer Goldstone modes governed by $K[V]$:
\begin{align}
  &  Z\approx \frac{1}{\mathcal{N}}  \int DU\, e^{- K[U]} \makebox[.5in]{,} 
K[U] \approx \gamma \vartheta^2   \sum_{p}{\rm Tr} (I - U(p) )  \;.
\label{gm}
\end{align}
This is nothing but the Wilson action for the $U(N)$ gauge field theory. The representation of a condensate of Abelian loops in terms of Abelian gauge Goldstone modes was done for the first time in Ref. \cite{Rey}. Here, relying on the Weingarten representation, this result was generalized to percolating oriented surfaces with $N$ possible labels at their vertices. Note that, when $\vartheta^2 > 0$ and $\lambda \to \infty$, the non-Gaussian integral with measure $dV$
in Eq. \eqref{Qmodel}  becomes a  group integral over $U(N)$ with Haar measure
$d\mu(U)$:
\begin{gather}
  \int  dV\, z^{-1}\,\exp{(- \mathcal{Q}(V)}) \, f(V) =   \int  d\mu(U) \, f(\vartheta U) \;.
\end{gather} 
Then, when $\lambda \to \infty$, the approximation in Eq. \eqref{gm} becomes exact. 

\begin{figure}
\centering
\begin{subfigure}{.5\textwidth}
  \centering
\includegraphics[scale=0.5]{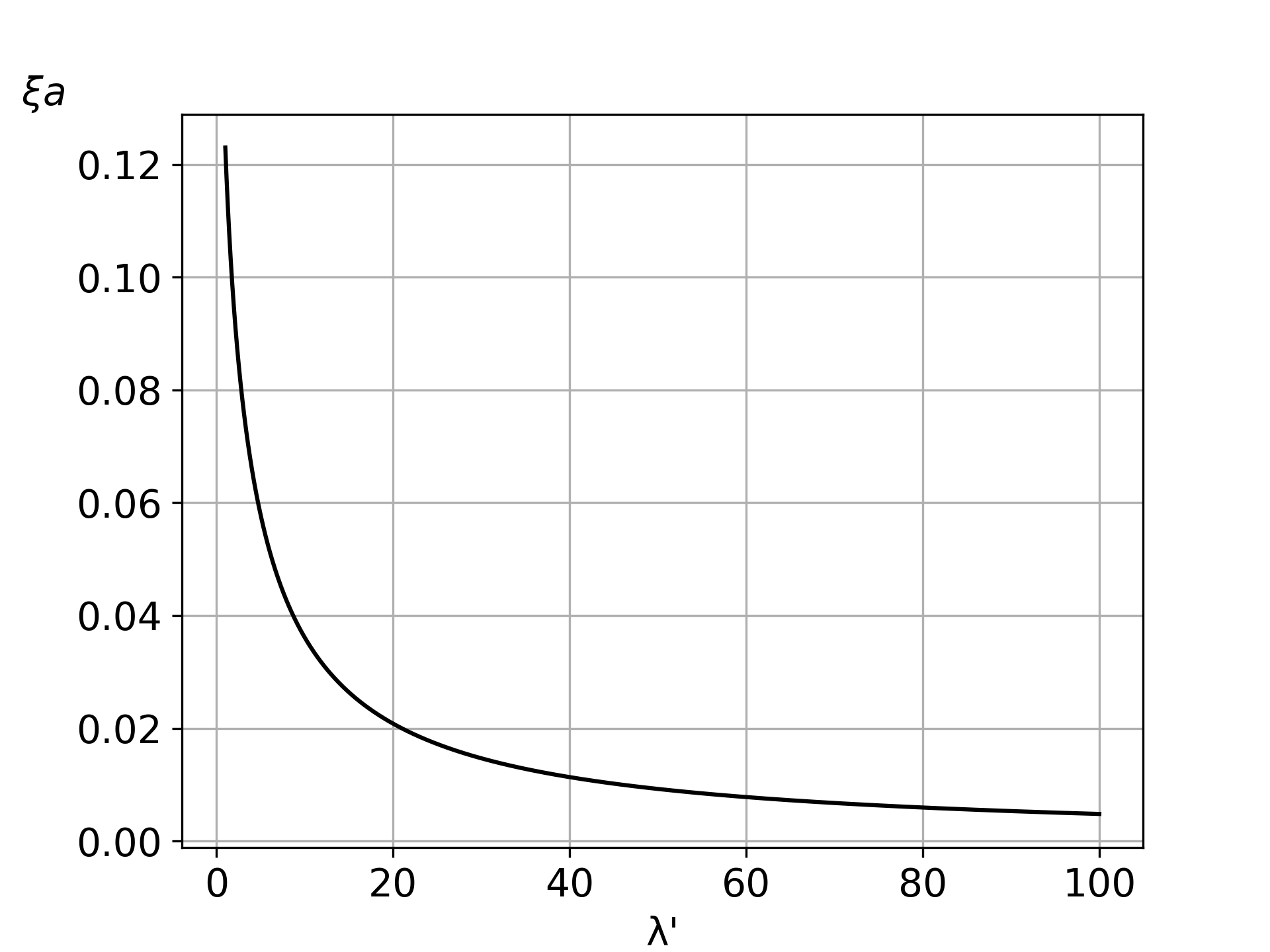}
\end{subfigure}%
\begin{subfigure}{.5\textwidth}
  \centering
\includegraphics[scale=0.5]{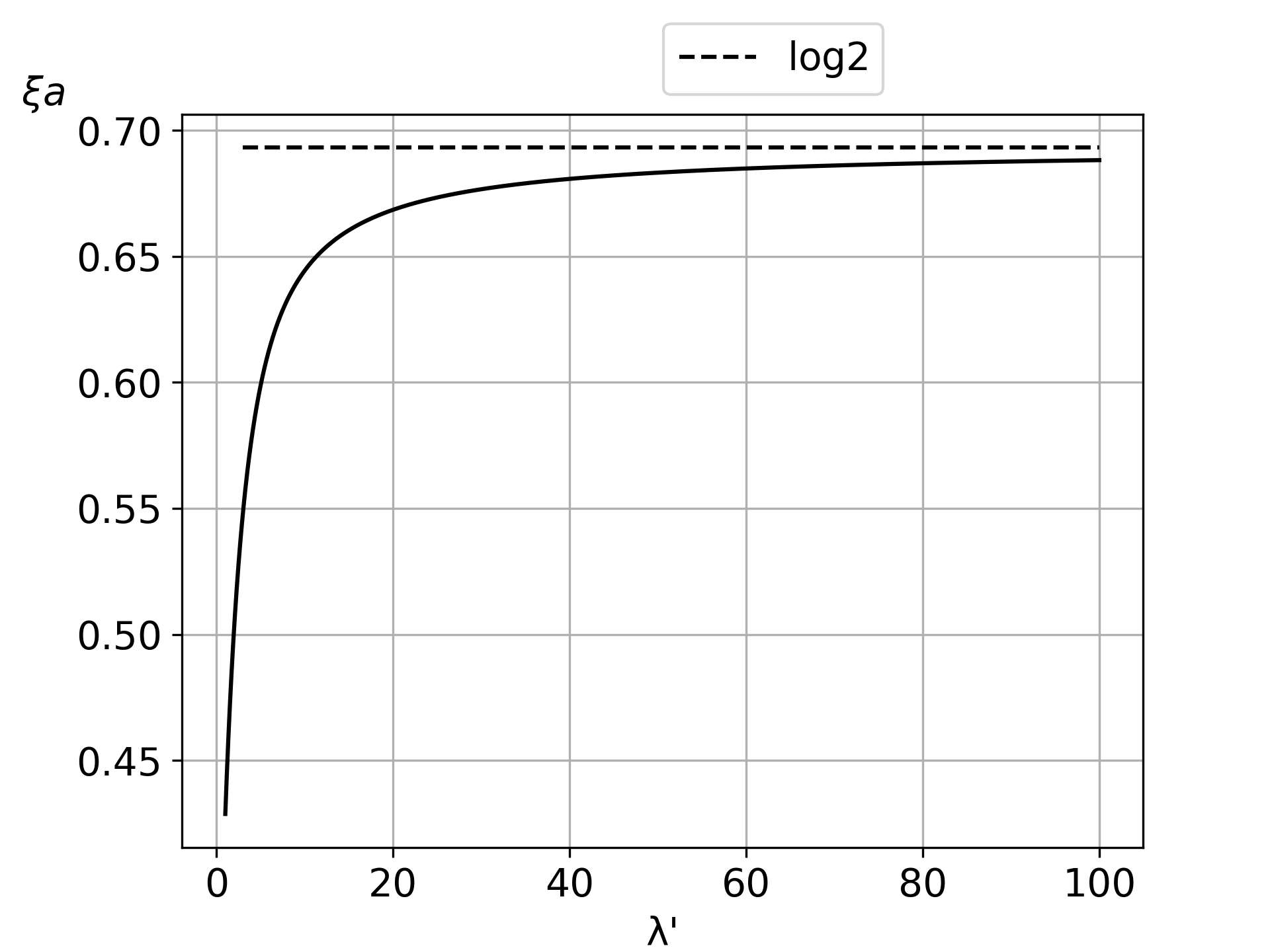}
\end{subfigure}
\caption{For $N=1$ and $\lambda' >0$ there is repulsion between surfaces: the interaction strength $\xi$ is positive in the normal $\vartheta^2=-1$ phase (left) and in the percolating $\vartheta^2=1$ phase (right). }
\label{fig:interaction}
\end{figure}

Regarding the interaction properties of the ensemble, we can look at the effect of doubly occupied links. For
$N=1$, we have 
\begin{align}
 \frac{\int dV \,  (V\bar{V})^2 \,  e^{-\mathcal{Q}(V)} }{\int dV \,  e^{-\mathcal{Q}(V)}}=  2\beta > 0 \makebox[.5in]{,}  C(p_1, \dots, p_{F})   
  =   2^{E_2}e^{-\mu A(\mathcal{S}) }  e^{- \xi L(\mathcal{C})} \makebox[.5in]{,} e^{-\xi a} = \frac{\beta}{\alpha^2}
     \;,
\end{align}
 where $L(\mathcal{C}) = a E_2$ is the length of the curve $\mathcal{C}$ where the surfaces are in contact. In Fig. \ref{fig:interaction}, we see that  the interaction between the surfaces is repulsive ($\xi > 0$) in the normal as well as in the percolating phase. Then, in the Abelian case, besides excluded volume effects, the quartic interaction also provides some stiffness. In particular, a strong repulsion prevents a surface from bending at angles of $180$ degrees, which tends to prevent crumpling. For a general analysis of the importance 
 of stiffness to stabilize random surfaces, see Ref. \cite{ambjorn}. In the non-Abelian case ($N \geq 2$), interpreting the effect of the quartic term in Eq. \eqref{Qmodel} on the ensemble is a difficult task. For instance, when considering the four-plaquette array in Fig. \ref{fig7:a}, the gluings in Figs. \ref{fig7:b} and \ref{fig7:c}  gain a factor $\beta_a$ (see Appendix \ref{doc}). In addition, two new terms with a factor $\beta_b$ are generated. They are given by a holonomy along the path $\partial (p \cup  \bar{p}) $ composed with $ \partial (q \cup \bar{q})$ and the path $\partial (p \cup  \bar{q}) $ composed with $ \partial (q \cup \bar{p})$, which define the border of new surface elements.
In particular, for a surface which along $\mathcal{C}$ is formed by a given sequence of $m$ ($n$) surface elements of type $a$ ($b$), the contribution is ($E_1 + 2E_2 = 2F$, $E_2 = m+n$)
\begin{align} 
   & N^{V(\mathcal{S})}e^{-\mu A(\mathcal{S})}  \left( \frac{\beta_a}{\alpha^2} \right)^m \left( \frac{\beta_b}{\alpha^2} \right)^n   \;. 
\end{align} 
Then, new combinations are generated with respect to the noninteracting Weingarten model, which increase for higher occupation. Over the years, numerous efforts have been made to interpret non-Abelian lattice models in terms of surfaces. Historically, these attempts were mainly concerned with determining a reference point where fundamental strings are free. 
For this reason, the focus was on implementing the 't Hooft large $N$-limit \cite{largeN} in lattice theories directly associated to the non-Abelian interactions. For instance, a thorough analysis aimed at establishing a link between the large-$N$ limit of $U(N)$ lattice gauge theory and strings was given in Refs.
 \cite{kaza}-\cite{kostov}. Also, a matrix model formed by the terms in Eq. \eqref{Qmodel}, nonlocal Coulomb interactions, and additional quartic terms, was considered in the large-$N$ limit to describe free Nambu-Goto strings 
\cite{stiffness-weingarten} and to study bound states in pure YM theory  \cite{bardeen}. In our work, the matrix model is also aimed at describing the strong interactions, but through the modeling of percolating center-vortex worldsurfaces. The ensemble interpretation can be given at the end, as its behavior will be essentially Abelianized due to the inclusion of chains and other relevant correlations observed in $SU(N)$ YM theory at finite $N$. Confining strings will also appear later, as topological solitonic flux tubes in the effective field descriptions.


\section{Ensembles of loops in the lattice}
\label{eloops}

In Ref. \cite{mixed},  the ensemble of holonomies along the monopole worldlines was treated as the continuum limit of a polymer growth process with stiffness.
In the next sections, we will need an ensemble representation in the same language used for surfaces. Then, let us briefly discuss the Weingarten representation of loops (for the Abelian case, see \cite{kawai1981}). Similarly to Eqs. \eqref{quarticmodel}, \eqref{Qmodel}, consider the partition function  
\begin{align}
    Z_m [H]= \frac{1}{\tilde{\mathcal{N}}}\int D\zeta\, \exp\left( \sum_{l} \tilde{V}(l)  -\tilde{Q}[\zeta]\right)\makebox[.5in]{,}
\tilde{V}(l) = \zeta^\dagger(x) H(x,y)\zeta(y) \;, 
\label{Zmo}
\end{align}
\begin{align}
\tilde{Q}[\zeta] =   \sum_x \tilde{\mathcal{Q}}(\zeta(x)) 
\makebox[.5in]{,}  \tilde{\mathcal{Q}}(\zeta)  =   \tilde{\eta} \,  \zeta^\dagger \zeta +
\tilde{\lambda}\,  (\zeta^\dagger \zeta)^2 \makebox[.5in]{,} \tilde{\mathcal{N}} =  \int D\zeta\, e^{- \tilde{Q}[\zeta]}\;, 
\end{align}
\begin{align}
 D\zeta =\prod\limits_{x} d\zeta(x)
 \makebox[.5in]{,} d\zeta(x) = 
 \prod\limits_{A}\pi^{-1}d[{\rm Re}\, \zeta_{A}(x)]d[{\rm Im}\, \zeta_{A}(x)]  \;,  \label{monopolenormalization}
\end{align}
where the sum is over the oriented links $l=(x,y)$ and the source satisfies $H(y,x) = H(x,y)^\dagger$. The tuple $\zeta$ is defined at the sites and is formed by $\mathcal{D}$ complex components, while $R(x,y)$ is a $\mathcal{D}\times \mathcal{D}$ matrix. In the $\tilde{V}$-expansion,  the contribution of sites with single occupation is distributed  on loops $\mathcal{C}_1,\mathcal{C}_2, \dots$, and is proportional to products of loop-variables
\begin{gather}
{\rm Tr}\, 	H(x_1,x_2)\dots H(x_{n},x_1) \;,
\label{prodH}
\end{gather}
where $(x_1,x_2),\dots , (x_{n},x_1)$ are the $n$ links forming $\mathcal{C}$. For example, if
\begin{gather}
	H(x,y) = \tilde{\gamma}\, R(x,y) \makebox[.5in]{,} R^\dagger(x,y) R(x,y) = I_{\mathcal D} \;,
	\label{fH}
\end{gather}
this contribution is given in terms of group holonomies $\Gamma_R(\mathcal{C})$:
\begin{gather}
 e^{-\tilde{\mu} (L(\mathcal{C}_1)+ L(\mathcal{C}_2) + \dots)} \, {\rm Tr}\, \Gamma_R(\mathcal{C}_1)\,{\rm Tr}\, \Gamma_R(\mathcal{C}_2) \dots \makebox[.5in]{,} 
\Gamma_R(\mathcal{C}) = R(x_1,x_2)\dots  R(x_{n},x_1)  \;, \label{R-holo} \\ \tilde{\gamma}\tilde{\alpha}= e^{-\tilde{\mu} a} \makebox[.5in]{,} L(\mathcal{C})=n a
 \makebox[.5in]{,} \tilde{\alpha} = \frac{1}{\mathcal{D}} \int d\zeta\, (\zeta^\dagger \zeta)  
\, e^{- \tilde{\mathcal{Q}}(\zeta)} \;.  \label{def-alpha}
\end{gather}
In this case, we can rewrite
\begin{gather}
    -  \sum_{l} \tilde{V}(l) +   \tilde{Q}[\zeta] =    \tilde{\gamma}\sum_{x,\mu}  (\Delta_\mu \zeta)^\dagger \Delta_\mu \zeta +  \sum_x
    \left( a^2 m^2  \zeta^\dagger \zeta 
    + \tilde{\lambda}\,  (\zeta^\dagger \zeta )^2 \right)
    \;, \nonumber \\
    \Delta_\mu \zeta =  R(x, x+ \mu) \zeta(x+ \mu) - \zeta(x) \makebox[.5in]{,}  \tilde{\eta} =   2d\tilde{\gamma} + a^2 m^2 \;,
    \label{Rg}
\end{gather} 
where $d$ is the spacetime dimension ($d=4$ in our case).   In general, $m^2$ controls whether the model displays SSB or not.  As is well-known, when the tension $\tilde{\mu}$ goes below a critical value $\tilde{\mu}_{\rm c} > 0$, a condensate is formed ($m^2 < 0$). That is, bosons with sufficiently small mass and repulsive interactions condense. The condensation of loops spanning surfaces, discussed in the previous section, is the higher dimensional version of this property. Furthermore, if we redefine $ \zeta(x)\to a \, \zeta(x)$, Eq. \eqref{Rg} is the discretized action for the complex scalar $\zeta$: 
\begin{align}
    \int d^4x\, \left(  ( \tilde{\gamma}\, (D_\mu \zeta)^\dagger D_\mu \zeta   + m^2  \zeta^\dagger \zeta  + \tilde{\lambda}\,  (\zeta^\dagger \zeta))^2 \right) \;. 
\end{align}

\section{Abelian-projected ensembles} 
\label{ape}

In what follows, we shall obtain the 4D partition function for the ensemble represented by the vacuum wavefunctional reviewed in Sec. \ref{wavefun}.

\subsection{Abelian-projected center vortices}

 To arrive at the wavefunctional $\Psi(A)$ in Eq. \eqref{wavef}, we initially considered 
 elementary center-vortex loops (see Sec. \ref{percolate}), that is, their center charge is $\pm 1$. As already discussed, there are $N$ Cartan species labeled by the (magnetic) weights of the defining representation $\beta_i$. 
 In this Hamiltonian approach, the wavefunctional  
was simply localized on the Abelian-projected gauge fields $a(\{ \gamma\})$ for the networks $\{ \gamma\}$   (cf. Eq. \eqref{sp2}). At the level of the 4D partition function, we shall initially consider an Abelian-projected ensemble of thick collimated configurations. This means that the Cartan gauge fields are such that their flux is localized around  center-vortex worldsurfaces. At distances larger than the typical transverse localization scale, away from the distribution $\{ \mathcal{S}\}$ of worldsurfaces,  the field-strength will be considered to be essentially zero. The configuration $\{ \mathcal{S}\}$ also includes a distribution of defining weights. As the different species $\beta_i$ are physically equivalent, the ensemble must be invariant under their permutations (Weyl transformations). On a fixed-time slice, the networks of worldsurfaces and weights  $\{ \mathcal{S} \}$ produce the networks of lines and weights  $\{ \gamma \}$ used to derive the effective representation of $\Psi(A)$.  They include the possibility of $N$-matching and nonoriented collimated chains. Away from $\{ \mathcal{S} \}$,  an oriented thick configuration $A_\mu^{\rm thick}$ is a superposition $a(\{\mathcal{S}\})$ of the gauge fields $a_\beta(\mathcal{S})$, with spacetime components \cite{cv-3}
\begin{gather}
a_\beta(\mathcal{S})|_\mu = 4\pi\beta\cdot T \,(-\partial^2)^{-1}\,\partial_\nu s_{\mu \nu} (\mathcal{S})  = 2\pi \beta\cdot T \int_{\mathcal{S}}  d^2 \tilde{\sigma}_{\mu \nu} \, \partial_\nu D(x-\bar{x}) \makebox[.5in]{,}  \partial^2 = \partial_\mu \partial_\mu\;,
\label{asigma}  \\
s_{\mu \nu}(\mathcal{S}) = \frac{1}{2}\int_{\mathcal{S}} d^2\tilde{\sigma}_{\mu\nu}\, \delta(x-\bar{x}) \makebox[.5in]{,} d^2 \tilde{\sigma}_{\mu \nu} = 
du dv\,  \frac{\partial \bar{x}_\rho}{\partial u} \frac{\partial \bar{x}_\sigma}{\partial v} \epsilon_{\mu \nu \rho \sigma} \label{source} \;,
\end{gather}
where $\bar{x} = \bar{x}(u,v)$ is a parametrization of the worldsurface $\mathcal{S}$ and $D(x-y)$ is the Green function: $-\partial^2 D(x-y) = \delta(x-y)$. The gauge field $a_\beta(\mathcal{S})$ generalizes $a_\beta(\gamma)$ in Eq. \eqref{a3d} to 4D Euclidean spacetime. For closed $\mathcal{S}$, it is equivalent to  $\beta\cdot T \, \partial_\mu \chi$ in Eq. \eqref{simple-vortex}.
Here, the orientation refers to the Lie algebra, that is, an oriented collimated configuration has no monopole worldlines carrying Cartan flux. This also applies to configurations with $N$-matching. In particular, the $N$-matching example in Eq. \eqref{Nex} corresponds to
 \begin{gather}
a(\{ \mathcal{S}\}) =  a_{\beta_1}(\mathcal{S}_1)+ \dots + a_{\beta_N}(\mathcal{S}_N)   \;,
\label{Nmat} 
\end{gather}
\begin{gather}
\beta_1 + \dots + \beta_N = 0  \makebox[.5in]{,} \partial \mathcal{S}_1 = \dots = \partial \mathcal{S}_N 
\label{can1}
   \end{gather}
where the $\mathcal{S}_i$'s are open surfaces with the same boundaries. 
 The gauge field description can be easily generalized to the nonoriented case. However, besides the physical center vortex guiding-centers $\{ \mathcal{S}\}$, the Abelian-projected variables gain unobservable Dirac worldsheets $\mathcal{U}$, 
 \begin{gather}
a(\{ \mathcal{S}\})  \to a(\{ \mathcal{S}\}) + a(\{ \mathcal{U}\})  \;. 
\label{non}
\end{gather}
 In particular, the nonoriented chain in Eq.  \eqref{epsilon} is generalized to  
\begin{gather}
 a(\{ \mathcal{S}\}) = a_{\beta}(\mathcal{S})+a_{\beta'} (\mathcal{S}')\makebox[.5in]{,} a(\{ \mathcal{U}\}) = a_\mathscr{E}(\mathcal{U}) 
  \;,
\end{gather}
\begin{gather}
  \mathscr{E} = \beta - \beta' \makebox[.5in]{,}  \partial\mathcal{U} =\partial \mathcal{S}'  = - \partial \mathcal{S}   \;.
  \label{can2}
\end{gather}
where the only physical information about $\mathcal{U}$ is on its boundary $\partial\mathcal{U}$, where the monopole worldline carrying the adjoint magnetic weight $\beta -\beta'$ is placed. We underline that, for thick configurations, the terms $a_{\beta}(\mathcal{S}) +a_{\beta}(\mathcal{S}')$ in the 
Abelian-projected nonoriented variable is only attained away from $\mathcal{S}$ and $\mathcal{S}'$, on the other hand, $a_\mathscr{E}(\mathcal{U})$ is always present. All these examples correspond to the Abelianized form of the gauge fields $ {\rm Ad}(a_\mu)$ in Eq. \eqref{Adj}, obtained from the singular mappings $S \in SU(N)$. Therefore, the worldsurface components in $\{ \mathcal{S} \}$ always form closed objects. 

Now, let us examine the flux collimation properties and their implications for modeling the average over an ensemble of thick configurations $A_\mu^{\rm thick}$.\footnote{The Wilson loop average over $A_\mu$ with a measure for $N-1$ compact QED(4) gauge fields, which corresponds to a monopole-only scenario, will be discussed in Appendix \ref{mon-only} for quarks in the defining representation.} For heavy quark probes in an irreducible  representation ${\rm D}(\cdot)$ with dimensionality $\mathscr{D}$,  the Wilson loop for a Cartan gauge field $A_\mu$ is:
\begin{gather}  
W_{\mathcal{C}_{\rm e}}[A]   =  \frac{1}{\mathscr{D}} \, {\rm Tr}\, {\rm D}  \left(  e^{i \oint_{\mathcal{C}_{\rm e}} dx_\mu\,  A_{\mu}  }  \right)  =  \frac{1}{\mathscr{D}} \,  \sum_{\omega_{\rm D}}  e^{i  (\Phi , \,
\omega_{\rm D} \cdot T)  } \nonumber \\ 
\Phi =  \oint_{\mathcal{C}_{\rm e}} dx_\mu\,  A_{\mu}  = \int d^4x\, \tilde{F}_{\mu \nu} s_{\mu \nu}(\mathcal{S}_{\rm e}) \makebox[.5in]{,}  \tilde{F}_{\mu \nu} = \frac{1}{2} \epsilon_{\mu \nu \rho \sigma} F_{\rho \sigma} \;,
\label{wloope}
\end{gather} 
where $\omega_{\rm D}$ are the weights of $D(\cdot)$ and ${\mathcal S}_{\rm e}$ is any surface whose boundary is ${\mathcal C}_{\rm e}$. Note that the source $s_{\mu \nu}(\mathcal{S}_{\rm e})$ (cf. Eq. \eqref{source}) implements Stokes' theorem. 
The scalar product $(X,Y)$ between Lie algebra elements is defined  such that $(T_A,T_B) = \delta_{AB}$.
For collimated fluxes, the detail about how they are distributed  around $\{ \mathcal{S} \}$ is not relevant to compute the Wilson loop at large distances. Indeed, let us consider the gauge field for  
\begin{align}
A^{\rm thick}_\mu = a^{\rm thick}_\mu + a_\mu(\{\mathcal{U}\})  \;.
\end{align}  
 If all the flux passes either through or outside $\mathcal{S}_{\rm e}$, then on $\mathcal{C}_{\rm e}$, $a^{\rm thick}_\mu $ can be replaced by $a_\mu(\{\mathcal{S}\})$, so that the enclosed flux $\Phi$ is  
\begin{align}  
  \Phi = \oint_{\mathcal{C}_{\rm e}} dx_\mu \Big( a_\mu(\{\mathcal{S}\}) + a_\mu(\{\mathcal{U}\}) \Big)  = \int_{\mathcal{S}_{\rm e}}  d^4x\,    \tilde{f}_{\mu \nu} s_{\mu \nu}(\mathcal{S}_{\rm e})  \;,
\end{align}
where we used Stokes' theorem once again and defined
\begin{gather}
 \tilde{f}_{\mu \nu} =  \tilde{f}_{\mu \nu}(\{\mathcal{S}\}) + \tilde{f}_{\mu \nu}(\{ \mathcal{U}\})   \;,
\label{Ft1}
\end{gather} 
which is a superposition of the dual field-strengths
\begin{eqnarray}
 &&\tilde{f}_\beta(\mathcal{S})|_{\mu \nu} 
 =  \epsilon_{\mu \nu \rho \sigma} \partial_\rho a_\beta (\mathcal{S})|_\sigma \nonumber \\
 &&= 2\pi\beta\cdot T \int_{\mathcal{S}}  d^2\sigma_{\mu\nu}\, \delta(x-\bar{x})  + 2\pi \beta\cdot T\left(\int_{\partial\mathcal{S}} dx_\nu \,\partial_\mu D(x-\bar{x})-\int_{\partial\mathcal{S}} dx_\mu \,\partial_\nu D(x-\bar{x})\right)  \;.
\end{eqnarray} 
The first term is concentrated on $\mathcal{S}$, while  for a closed $\mathcal{S}$ the second term vanishes ($\partial \mathcal{S} = 0$). However, for an open component in the center-vortex network, the second term spreads in 4D Euclidean spacetime from $\partial \mathcal{S}$. This is similar to what happens with $a_\beta(\gamma)$ in Eq. \eqref{a3d} when computing $f_\beta(\gamma) = \nabla \times a_\beta(\gamma)$, which contains a term localized on $j(\gamma)$ plus fluxes isotropically distributed from their endpoints, where the density $\nabla \cdot j(\gamma)$ is concentrated. The important point is that for arrays with $N$-matching and nonoriented chains, because of the flux distribution at the boundaries of its components (see, e.g. Eqs. \eqref{can1} and \eqref{can2}), the noncollimated terms are canceled. This only leaves the collimated components
\begin{gather}
 \tilde{f}_{\mu \nu} = \sum 2\pi\beta\cdot T \int_{\mathcal{S}}  d^2\sigma_{\mu\nu}\, \delta(x-\bar{x}) + \sum 2\pi\mathscr{E} \cdot T \int_{\mathcal{U}}  d^2\sigma_{\mu\nu}\, \delta(x-\bar{x})   \;,
\label{Ft2}
\end{gather} 
where the sums in the first and second terms are taken over $(\beta,\mathcal{S}) \in \{ \mathcal{S} \}$ and $(\mathscr{E}, \mathcal{U}) \in \{ \mathcal{U} \}$, where $\mathscr{E}$ is the difference of the defining magnetic weights carried by the pair of center vortices attached to a monopole worldline.   Therefore, if all the flux passes through or bypasses $\mathcal{S}_{\rm e}$, it can be simply written in terms of the intersection number between $\mathcal{S}$ and $\mathcal{S}_{\rm e}$ 
\begin{align}  
   I(\mathcal{S},\mathcal{S}_{\rm e}) =  \frac{1}{2}  \int_{\mathcal{S}}  d^2\sigma_{\mu\nu} \int_{\mathcal{S}_{\rm e}} d^2\tilde{\sigma}_{\mu\nu}\,  \delta(\bar{x} -\bar{y}) =
    \int_{\mathcal{S}}  d^2\sigma_{\mu\nu} \,  s_{\mu \nu}(\mathcal{S}_{\rm e})  \;,
\end{align}
where $\bar{y}$ is a parametrization of $\mathcal{S}_{\rm e}$. In effect, from Eqs. \eqref{source} and \eqref{Ft2}, we get
\begin{align}
    \Phi = \Phi(\{\mathcal{S}\}) + \Phi (\{\mathcal{U}\})  = \sum 2\pi \beta \cdot T  \, I(\mathcal{S},\mathcal{S}_{\rm e}) +  \sum 2\pi \mathscr{E} \cdot T  \, I(\mathcal{U},\mathcal{S}_{\rm e})   \;.
    \label{enclosed}
\end{align}  
In general, for a given realization in the ensemble, some fluxes will partially pass through $\mathcal{S}_{\rm e}$. However, for asymptotically large $\mathcal{C}_{\rm e}$ relative to the transverse localization scale of thick center vortices, the effect of these occurrences on $\langle W_{\mathcal{C}_{\rm e}} \rangle$ can only scale with the perimeter of $\mathcal{C}_{\rm e}$. In this regime, we shall disregard them and consider Eq. \eqref{enclosed} to compute the Wilson loop:
\begin{gather}
 W_{\mathcal{C}_{\rm e}}  =  \frac{1}{\mathscr{D}} \, {\rm Tr}\, {\rm D}  \left(  e^{i \Phi} \right)  =  e^{-i\frac{2\pi k}{N} \sum_{\mathcal{S}} I( \mathcal{S},\mathcal{S}_{\rm e})} \;,
\end{gather}
where we used Eq. \eqref{DPa2}, which in particular gives $e^{i 2\pi  \,  \beta_i \cdot T }= e^{-i \frac{2\pi }
 		{N} } I $ and $e^{i 2\pi  \,  \mathscr{E} \cdot T } = I $.   
Of course, the Dirac worldsurfaces have no effect on  observables. The particular defining weights attributed to each component in $\{ \mathcal{S} \}$ are also erased. In this approximation, the only information left in $W_{\mathcal{C}_{\rm e}}$ is their (common) associated center-element and the intersection numbers between the physical center-vortex worldsurfaces $\mathcal{S}$ and $\mathcal{S}_{\rm e}$. However, from a phenomenological perspective, when modeling the ensemble average $\langle W_{\mathcal{C}_{\rm e}} \rangle$, we must 
attribute weights $\psi_{\{\mathcal{S}\}}$ that not only characterize properties of elementary center vortices (such as tension and stiffness), but also parametrize the importance of different interactions, $N$-matching, and nonoriented components:
\begin{gather} 
 \langle W_{\mathcal{C}_{\rm e}} \rangle  =	\sum_{\{ \mathcal{S} \} } \psi_{\{ \mathcal{S}\}} \,  e^{-i\frac{2\pi k}{N} \sum_{\mathcal{S}} I( \mathcal{S},\mathcal{S}_{\rm e})}   \;. 
 	\label{DP}
 \end{gather} 
    This average can be regularized in the lattice, along with the main properties of center-vortex ``matter''.  The key elements for this task were established in the Weingarten framework, in Secs. \ref{wsur}, \ref{stabilized}, and \ref{eloops}. There, we showed that the Gaussian matrix model for the link-variables $V(x,y) \in \mathbb{C}^{N \times N}$ (cf. Eq. \eqref{matrixmodel}) generates weights for noninteracting closed worldsurfaces only characterized by a tension $\mu_0$ and $N$ possible labels at their vertices (cf. \eqref{sum-colors}).   Just one global label $\beta$ simply corresponds to $N=1$:
\[
 Z_0 =\int DV\, \exp\Big(\gamma\sum_{p} V(p)-Q_0[V]\Big) = \sum_{\{ \mathcal{S} \} } e^{- \mu_0  \sum _{\mathcal{S}}A(\mathcal{S})} \makebox[.5in]{,} V(x,y) \in \mathbb{C}  \;.
\]
With this measure, and taking $\mathcal{C}_{\rm e}$ formed by links in a lattice that is dual to the one used in the Weingarten representation,  the regularized average in Eq. \eqref{DP} would be
\begin{gather} 
  Z_0[b] = \int DV\, \exp\Big(\gamma\sum_{p} e^{i b(p)} V(p)-Q_0[V]\Big) =\sum_{\{ \mathcal{S} \} }e^{- \mu_0  \sum _{\mathcal{S}}A(\mathcal{S})} \,  e^{-i\frac{2\pi k}{N} \sum_{\mathcal{S}} I( \mathcal{S},\mathcal{S}_{\rm e})}   \;, 
 \end{gather} 
 where $b(p) = (2\pi k/N) s(p)$ is nontrivial only if $p$ intersects $\mathcal{S}_{\rm e}$,  $\partial \mathcal{S}_{\rm e} =\mathcal{C}_{\rm e}$. In such cases, $s(p)$ equals $+1$ ($-1$) depending on whether the relative orientation between $p$ and $\mathcal{S}_{\rm e}$ at the intersection point is positive (negative). In effect, the $\gamma$-expansion of $Z_0[b]$ continues generating contributions from plaquettes distributed over closed surfaces $\mathcal{S}$ (like in $Z_0$), while the frustration $e^{ib(p)}$ produces a center-element every time $\mathcal{S}$ intersects $\mathcal{S}_{\rm e}$. Note that the generation of a center-element average as due to a lattice frustration was proposed in Ref. \cite{mixed} (see Sec. \ref{zf}).  As already discussed in Sec. \ref{stabilized}, this ensemble can be stabilized by  a quartic term that provides excluded volume effects and stiffness at 180 degrees.
In the Abelian-projected setting, it is clear that the partition function can be constructed in terms of $N$ copies $V_i(x,y) \in \mathbb{C} $, $i=1, \dots , N$ that generate surfaces associated with the global magnetic charges $\beta_i$.  This association will be established by introducing new interactions and scalar fields into the Weingarten model, designed to produce the main correlations among the generated surfaces. Meanwhile, the average of the Wilson loop over the $N$ types of worldsheets is given by 
\begin{align}
\big(Z[b]\big)^N=   \frac{1}{\mathcal{M}} \int  \int  \prod_i  DV_i\, \exp\Big( \gamma\sum_{p}{\rm Tr} \big( e^{i b(p)} V(p) \big) - Q[V]\big)\Big)  \;,
 \label{prodi}
\end{align}
where we introduced a diagonal $N \times N$ matrix $V$:
\begin{gather}
V = \left(
\begin{array}{cccc}
V_1 & 0 & \cdots & 0 \\
0 & V_2 & \cdots & 0 \\
\vdots & \vdots & \ddots & \vdots \\
0 & 0 & \cdots & V_N
\end{array}
\right)
\;.
\label{Vmat}
\end{gather}
Note that $Q[V]$ in Eq. \eqref{prodi} is given by Eq. 
\eqref{Qmodel} but applied to this $N \times N$ diagonal matrix. Up to this point, the  model displays a $U(1)^N$ local symmetry
\begin{gather}
	V(x,y) \to U(x) V(x,y) U^\dagger(y)  \makebox[.5in]{,}
	U(x)  \in U(1)^N  \;.
	\label{n-2sym}
\end{gather}
 Now, similarly to the wavefunctional in Eq. \eqref{wavef},
the natural $N$-matching among vortices discussed in Sec. \ref{N-matching} can be simply implemented by 
\begin{gather}
 Z_{\rm c.v.}[b] =   \frac{1}{\mathcal{M}} \int    DV\, \exp 
 \left( -W_{\rm c.v.}[V] \right)  \makebox[.5in]{,} DV = DV_1 \dots DV_N  \;,  \nonumber \\
W_{\rm c.v.}[V] = -\gamma\sum_{p}{\rm Tr} \big( e^{i b(p)} V(p) \big) + Q[V] - \sum\limits_{\{x,y\}} \xi \big(\det V(x,y) + \det V(y,x)\big) \;.
\label{wN-m}
\end{gather}
Indeed, upon expanding in powers of $\xi$ and $\gamma$, the only nontrivial contributions arise from terms where all variables are paired. The new term $\xi \, V_1(x,y) \dots V_N (x,y)$ allows for the possibility of having  $N$ open surfaces carrying the weights $\beta_1,\dots,\beta_N$  ($\beta_1+\dots+\beta_N=0$) that meet at links $(x,y)$ that form loops.   With this $N$-matching property, the local symmetry is reduced to $U(1)^{N-1}$: 
\begin{gather}
  V(x,y) \to U(x) V(x,y) U^\dagger(y)  \makebox[.5in]{,}
    U(x)= e^{i \xi(x) \cdot T}  \;.
\end{gather}

\subsection{Abelian-projected chains}

\label{abe-chains}

Here, in the context of the Weingarten representation,   we shall include the chains discussed in Sec. \ref{cha}. 
This can be easily done by means of a factor  $Z_{\rm m}[H]$ that generates loops (cf. Eq. \eqref{Zmo}) in the integrand of Eq. \eqref{wN-m}, which generates surfaces:
\begin{gather}
Z_{\rm mix}[b]   \propto \int    DV D\phi\, \exp 
 \left( -W_{\rm mix}[V, \phi]\right) \makebox[.5in]{,}
W_{\rm mix}[V, \phi] =W_{\rm c.v.}[V] - \sum_{l} \tilde{V}(l) +\tilde{Q}[\phi] \;, \nonumber \\
\tilde{V}(l) = \sum_{\alpha}\bar{\phi}_{\alpha}(x) H_\alpha(x,y) \phi_{\alpha}(y) \makebox[.5in]{,} H_\alpha(x,y) = \bar{V}_j(x,y)V_i(x,y)  \;, \nonumber \\
\tilde{Q}[\phi] =   \sum_x \sum_{\alpha} \tilde{\mathcal{Q}}(\phi_{\alpha}(x)) 
\makebox[.5in]{,}  \tilde{\mathcal{Q}}(\phi)  =   \tilde{\eta} \,  \bar{\phi} \phi +
\tilde{\lambda}\,  (\bar{\phi} \phi)^2\;.
\label{abz}
\end{gather} 
Here, $\alpha$ is a shorthand notation for the pair of indices $ij$, $i<j$.  Note that when expanding in powers of $\tilde{V}$, products of loop variables  (cf. Eq. \eqref{prodH})
\begin{gather}
	H_\alpha(x_1,x_2)\dots H_\alpha(x_{n},x_1) = V_i(x_1,x_2)\dots  V_i(x_n,x_1)   V_j(x_1,x_n) \dots V_j(x_2,x_1) \;,  \nonumber
\end{gather} 
with the different $\alpha$ combinations, are generated. 
Thus, in the $\gamma$-expansion, 
 plaquettes of type $i$ can be glued to type $j$ along $\mathcal{C}$. That is, we gain chains where an open center-vortex worldsurface with global charge $\beta_i$ can be transformed into another with global charge $\beta_j$ at a monopole worldline.  Because of the frustration, center elements
 that depend on the linking number between chains and $\mathcal{C}_{\rm e}$ will also be present. This agrees with 
 the Wilson loop calculation, based on a non-Abelian gauge field labeled by $S$ in Eq. \eqref{Schain}, when the cores are completely linked by $\mathcal{C}_{\rm e}$.
 The mixed model continues to display a local $U(1)^{N-1}$ symmetry (cf. Eq. \eqref{n-2sym}), which also transforms the monopole fields
\begin{align}
     V_i(x,y) \to e^{i \xi(x)\cdot \omega_i}  \, V_i(x,y) \, 
     e^{-i \xi(y)\cdot \omega_i}  \makebox[.5in]{,} \phi_{\alpha}(x) \to e^{i \xi(x) \cdot \alpha_{ij} } \, \phi_{\alpha}(x)
     \makebox[.5in]{,} \alpha_{ij} = \omega_i - \omega_j  \;.
\end{align}
Since $\alpha_{ij}$ is a difference of defining weights, the label $\alpha$ can be identified with a (positive) weight of the adjoint representation of $\mathfrak{su}(N)$.

\subsection{Percolating phase}
\label{abelian-percolating}
Now, we recall that in the wavefunctional approach to the mixed ensemble, there is a physically interesting sequence of phase transitions.  Initially, we considered the formation of a center-vortex condensate, where $N$-matching forces the vortex field $\Phi$ in Eq. \eqref{wavef} to be in $U(1)^{N-1}$. This was followed by a softer transition where monopoles are turned on. Let us consider the same process in the language of the 4D partition function $Z_{\rm mix}[b]$. 
From the discussion in Sec. \ref{stabilized}, when center-vortex worldsurfaces percolate and have repulsive interactions 
($ \eta < 0$, $ \lambda' = \lambda -3\gamma > 0$),  $Q[V]$ in 
Eq. \eqref{wN-m} drives the matrix $V(x,y)$ in Eq. \eqref{Vmat} to be close to $\vartheta\, U(x,y)$, $U(x,y) \in U(1)^{N}$ (see the discussion that led to Eq. \eqref{gm}). In addition, for a large $\xi$, the determinant favors $N$-matching and forces $U(x,y)$ to be close to $ U(1)^{N-1}$:
\begin{gather}
	U(x,y) = e^{i \theta(x,y)\cdot T } \makebox[.5in]{,} U_i(x,y) = e^{i \theta(x,y)\cdot \omega_i }\;. 
	\label{Ucomp}
\end{gather}
 If this initial stage dominates, for $\alpha = \alpha_{ij}$, we obtain 
\begin{gather}
	H_\alpha(x,y) = \vartheta^2\, R_\alpha(x,y) \makebox[.5in]{,}   R_\alpha(x,y) = U_i(x,y) \bar{U}_j(x,y) = e^{i \theta(x,y) \cdot \alpha  } \;. \label{abelian-gaugefield-latt}
\end{gather}
Therefore, when center vortices percolate, we may use 
Eq. \eqref{Rg} to approximate the mixed ensemble model:
\begin{gather}
W_{\rm mix}[V, \phi] \approx  \gamma \vartheta^2   \sum_{p}{\rm Tr} \Big( I -  e^{i b(p)}  U(p) \Big) +  \sum_{x,\mu, \alpha} \vartheta^2(\Delta_\mu \phi_\alpha)^\dagger \Delta_\mu \phi_\alpha + \mathcal{U}(\phi) 
     \;, \nonumber \\
      \Delta_\mu \phi_\alpha =  R_\alpha(x, x+ \mu) \phi_\alpha(x+ \mu) - \phi_\alpha(x) \makebox[,5in]{,} \mathcal{U}(\phi)= \sum_{x,\alpha}  
    \big(a^2 m^2  |\phi_\alpha|^2 
    + \tilde{\lambda}\,  |\phi_\alpha|^4 \big)   \;.
     \label{Wmix}
\end{gather} 
As in previous works, besides the center-vortex condensate 
we shall also assume a monopole condensate ($m^2 < 0$). After disregarding boundary effects, the only information left about the initial quark representation ${\rm D}(\cdot)$ is $k$, so that asymptotic $N$-ality is explicitly implemented.  In particular, for adjoint quarks, there is no frustration at all and the Wilson loop average  gives just $1$ (no confining string). 

\subsection{Abelian projection and the confining flux tube}
\label{Abe-ft}

\begin{figure}
\centering
\includegraphics[scale=0.5]{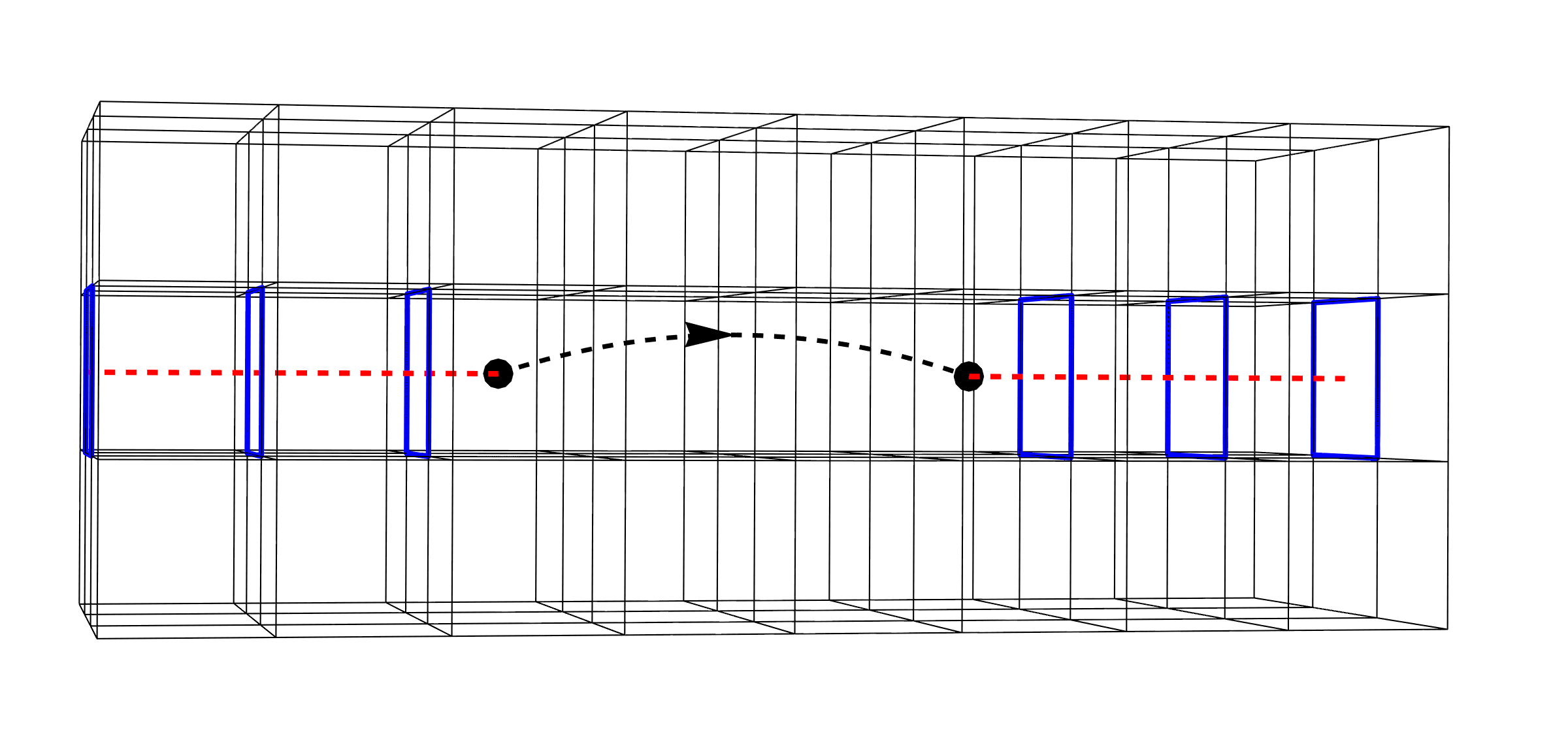}
\caption{At an intermediate time-slice ($x_0=0$), we see a quark-antiquark pair (black dots) that belongs to 
the Wilson loop $\mathcal{C}_{\rm e}$. The dotted red lines belong to  $S_{\rm e}$, while the black lines represent the observable guiding-centers of the flux tube.  The plaquettes $\tilde{p}$ that intersect $S_{\rm e}$ are in blue.   }
\label{fig:wloop}
\end{figure}

For nontrivial $k$, let us initially consider a Wilson loop along a large rectangular path $\mathcal{C}_{\rm e}$  in the $x_0, x_3$ plane. The frustration can be assigned to the plaquettes that intersect any surface $\mathcal{S}_{\rm e}$,  $\partial \mathcal{S}_{\rm e} =\mathcal{C}_{\rm e}$. For definiteness, let us consider the surface  $|x_0|  > T/2$, $|x_ 1| > L/2$, $x_2 =0$, $x_3 =0$. At an intermediate time-slice ($x_0 =0$), we have a cubic lattice in $\mathbb{R}^3$ (Fig. \ref{fig:wloop}). 
The cubes centered at $x_1 = \pm L/2$, $x_2=0$, $x_3 =0$ contain the quark and antiquark.  The partition function  $Z_{\rm mix}[b] $ will be dominated by the lowest action configurations of $W_{\rm mix}[V, \phi]$, which could be directly obtained from computer lattice calculations. To gain insight about  them, we note that at large distances from the quark-antiquark pair, they should be close to 
\begin{gather}
	U(p) = e^{-ib(p)} I \makebox[.5in]{,}   R_\alpha(x, y) \phi_\alpha(y) = \phi_\alpha(x)    
 \makebox[.5in]{,} \bar{\phi}_\alpha \phi_\alpha = v^2\;,
	\label{vacu}
\end{gather}
where $v^2=-a^2 m^2/(2\tilde{\lambda})$. This gives (cf. Eq. \eqref{abelian-gaugefield-latt}) 
\begin{gather}
 U(l) = e^{i \Delta(l) \cdot T } 
    \makebox[.5in]{,} e^{i\sum_{l\in \partial p} \Delta(l) \cdot T } =  e^{-ib(p)} I \makebox[.5in]{,}  e^{i \Delta(l) \cdot \alpha } \phi_\alpha (y) =   \phi_\alpha(x) \;,
    \label{delta-con}
\end{gather}
where $l=(x,y)$. 
When $a\to 0$, at plaquettes $\tilde{p}$ with nontrivial frustration and positive orientation relative to $\mathcal{S}_{\rm e}$, we must also have 
\begin{gather}
e^{i\sum_{l\in \partial  \tilde{p} } \Delta(l) \cdot T}= e^{-i\frac{2\pi k}{N} } I\;.
\label{conti}   
\end{gather} 
  This is because $b(p)$ does not scale as $a^2$, so its effect,  concentrated on a surface, would produce a divergent action. 
To solve these conditions, note that Eq. \eqref{DPa2} is valid for any representation ${\rm D}'(\cdot)$ of $SU(N)$ with $N$-ality $k$. 
 Using \[
 {\rm D}' \left(  e^{i 2\pi   \,  \beta_i \cdot T   } \right) = 
   e^{i 2\pi   \,  \beta_i \cdot  {\rm D}' (T)   } 
   \]
   and acting on any weight-vector of ${\rm D}'(\cdot)$ with weight $\omega'$, we have $e^{i 2\pi  \,  \beta_i \cdot \omega'} =e^{-i 2\pi k/N } $. Equivalently, 
\begin{gather}
	  e^{i 2\pi  \,  \beta' \cdot T}  =  e^{-i \frac{2\pi k}{N} }  I\makebox[.5in]{,}   \beta' = 2N \omega'\;. 
\end{gather}
On the other hand, the nontrivial plaquettes $\tilde{p}$ at $x_0=0$,  are those that intersect the $x_1$-axis for $|x_1|> L/2$. Then, as the continuum limit is approached, away from the quark-antiquark pair or on the links belonging to $\partial \tilde{p}$, the finite-action modes must be characterized by 
\begin{gather}
    \Delta(l) = \beta' \Delta_l \xi \;.
\label{Delta-defi}
\end{gather}
Here,  $\xi$ is a multivalued phase that changes by $2\pi$ when going around a path $\tilde{\gamma}$ formed by the red and black dotted lines in  Fig. \ref{fig:wloop}. This angle, defined in the continuum, is embedded in the lattice through its variation  $\Delta_l \xi$ along $l$, which is a well-defined link-variable. In this manner, $\sum_{l \in \partial p} \Delta(l) $ gives $2\pi \beta'$ or $0$, depending on whether $\partial p$ links  $\tilde{\gamma}$ or not, which implies the second condition in Eq. \eqref{delta-con}. Moreover, Eq. \eqref{Delta-defi} leads to well-defined monopole fields: starting from a site $x$ and transforming $\phi_{\alpha}(x)$ along $\partial \tilde{p}$, due to the third condition we return to $\phi_\alpha(x)$:
\begin{gather}
    e^{i 2\pi \beta' \cdot \alpha } \phi_\alpha(x) = e^{i 2\pi  \,  \beta' \cdot \omega_i} e^{-i 2\pi  \,  \beta' \cdot \omega_j} \phi_\alpha(x) = \phi_\alpha(x) \;.
\end{gather} 
Now, the key question is which configuration has the lowest action. Of course, a lower action will be obtained by considering a straight line between the quark and the antiquark. But what about the magnetic weight $\beta'$, with $N$-ality $k$, that must be used?  This can be addressed by considering the continuum limit of the lattice field equations. By using $\theta(x,y) = a \Lambda_\mu(x)$ in Eqs. \eqref{Ucomp}, \eqref{abelian-gaugefield-latt} and the redefinition $\phi_\alpha \to a \phi_\alpha$, ($v\to a v$) these equations become
\begin{gather}
D^2\phi_\alpha=m^2 \phi_\alpha + 2\tilde{\lambda}\phi_\alpha(\bar{\phi}_\alpha \phi_\alpha)\makebox[.5in]{,} D_\mu =\partial_\mu-i \Lambda_\mu \cdot \alpha \;,\nonumber\\
\partial_\mu F_{\mu\nu} = \gamma \vartheta^2\sum_\alpha\left(-i(\partial_\nu\bar{\phi}_\alpha\phi_\alpha-\bar{\phi}_\alpha \partial_\nu \phi_\alpha)+2 \Lambda_\nu \cdot \alpha\, \bar{\phi}_\alpha \phi_\alpha \right) \alpha \cdot T \makebox[.5in]{,}F_{\mu\nu}=\partial_\mu  \Lambda_\nu-\partial_\nu  \Lambda_\mu\;. 
\label{fe-cont}
\end{gather}
In addition, away from the quark-antiquark pair, the conditions in Eq. \eqref{delta-con} now read
\begin{align}
&
\Lambda_\mu \to \beta' \cdot T\, \partial_\mu\xi   \makebox[.5in]{,}
  \phi_\alpha(\rho ) \to v e^{i\beta'\cdot \alpha\, \xi }\;,
  \label{asy-cont}
\end{align}
which must also be the behavior close to the $x_1$-axis, around the pair of lines $|x_1| > + L/2$.  In other words, in the 4D Euclidean spacetime, the field-strength near $\mathcal{S}_{\rm e}$ must approach
\begin{align}
&
F_{\mu \nu}(\Lambda) \to J'_{\mu \nu} \makebox[.5in]{,}
J'_{\mu \nu} =  2\pi\beta'\cdot T s_{\mu\nu}(\mathcal{S}_{\rm e})\;,
\label{S-abe-a}
\end{align}
where $s_{\mu\nu}$ was defined in Eq. \eqref{source}. Moreover, the action must be computed with 
\begin{gather}
	 S(J')=\int d^4x \Big[ \frac{\gamma\vartheta^2}{4}(F_{\mu\nu}-J'_{\mu \nu})^2 +\sum_\alpha  D_\mu \bar{\phi}_\alpha D_\mu  \phi_\alpha+ \mathcal{U}(\phi)\Big] \;,
	\label{S-abe} \\ 
	\mathcal{U}(\phi) =\sum_\alpha\Big( m^2  |\phi_\alpha|^2+\tilde{\lambda} |\phi_\alpha|^4\Big) \;,
\end{gather}
evaluated on the particular saddle-point characterized by $\beta'$.\footnote{ The sum is over the positive roots ($\alpha$) of $\mathfrak{su}(N)$.} This corresponds to the continuum limit of $W_{\rm mix}[V, \phi] $.  In particular, note that when writing the frustration in terms of $\beta'$,  the lattice gauge-field sector in Eq. \eqref{Wmix},
\begin{align}
    &  \gamma \vartheta^2   \sum_{p}{\rm Tr} \Big( I - e^{i a^2 F_{\mu\nu} (p)-i  2\pi\beta'\cdot T s(p)}\Big) \makebox[.5in]{,} a F_{\mu\nu}(p) =  \Lambda_\nu(x+\mu)-\Lambda_\nu(x) -\Lambda_\mu(x+\nu)+\Lambda_\mu(x)\;,  \nonumber
\end{align}
 precisely generates the source $J'_{\mu \nu}$ in Eq. \eqref{S-abe}, which combined with the behavior in Eq. \eqref{S-abe-a} leads to a finite action density. For the lowest action, $\xi  $ is the polar angle $ \varphi$ with respect to the $x_1$-axis ($\tilde{\gamma}$ is a straight line).  In addition, for a given $N$-ality $k$, the lowest-action flux tube is expected to be associated with the $k$-Antisymmetric representation.  
 This can be explicitly shown at the BPS point, following the same analysis presented in Ref. \cite{O-S-J} for the non-Abelian case. When the distance $L$ is large, the solution on a plane between the quark and the anti-quark (see Fig. \ref{fig:wloop}) is approximately $x_1$-independent. The existence of the BPS point is evidenced by using the property
 \begin{gather}
     \sum_{\alpha >0} \alpha|_p \alpha|_q = \frac{\delta_{qp}}{2} 
     \label{alphasum}
 \end{gather}
and considering $\tilde{\lambda}$ such that $4\tilde{\lambda}\gamma\vartheta^2=1$, to write the energy density on the plane $x_1=0$ as a sum of squares, plus a topological charge:
\begin{align}
     &\sigma=\int d^2x\sum_\alpha\left(\gamma\vartheta^2B_\alpha^2+|D\phi_\alpha|^2+\frac{1}{4\gamma\vartheta^2}(\bar{\phi}_\alpha\phi_\alpha-v^2)^2\right)\makebox[.5in]{,}B_\alpha=B\cdot\alpha \;,\\&
     = \int d^2x\sum_\alpha  \left(|D_+\phi_\alpha|^2+\frac{1}{4\gamma\vartheta^2}\left(\bar{\phi}_\alpha\phi_\alpha-v^2+2\gamma\vartheta^2B_\alpha\right)^2+ v^2 B_\alpha  \right)\makebox[.5in]{,}D_+=D_2+iD_3\;. 
 \end{align}
 This leads to the BPS equations
\begin{align}
     D_+\phi_\alpha =0\makebox[.5in]{,}2\gamma\vartheta^2B_\alpha +\bar{\phi}_\alpha\phi_\alpha-v^2=0\;,
 \end{align}
which imply the second order Euler-Lagrange equations.
Then, at the BPS point, the tension is given by
 \begin{align}
\sigma= v^2 \sum_\alpha \int d^2x\, B_\alpha  =v^2\oint dx\cdot  A_\alpha \;.
 \end{align}
 Asymptotically, we have $A=\beta'\cdot T \, \nabla\varphi $, so that
 \begin{align}
     \sigma=2\pi v^2 \beta'\cdot\delta\makebox[.5in]{,} \delta=\sum_{\alpha>0}\alpha\;. \label{tension-bps}
 \end{align}
 Then, we can follow Ref. \cite{O-S-J} to show that the lowest $\sigma$ corresponds to 
 \begin{gather}
\beta' = \beta^{(k)}= 2N \omega^{(k)} 
\label{betak-a} \;,
\end{gather}  
 where $\omega^{(k)}$ is the highest weight of the $k$-Antisymmetric representation
 \begin{gather} 
\omega^{(k)} =  \omega_1 + \dots + \omega_k  
\label{betak-b}\;.
 \end{gather}
 The argument is based on initially considering the Young-Tableau for any representation with $N$-ality $k$, which fixes the  number of boxes as $k$ modulo $N$. When the number of boxes is precisely equal to $k$, the product $\beta'\cdot \delta$ turns out to be lower. Next,  the operation of lowering a box was shown to decrease this product. That is, the lowest tension corresponds to a Young-Tableau formed by a set of $k$ vertical boxes, which corresponds to the $k$-Antisymmetric representation. Before computing Eq. \eqref{tension-bps} as a function of $k$, let us determine the asymptotic scaling law for general parameters, not necessarily at the BPS point, assuming that $\beta^{(k)}$ produces the lowest tension.  Because of cylindrical symmetry, Eqs. \eqref{fe-cont} and \eqref{asy-cont} are solved with 
 \begin{align}	\Lambda_\mu=a(\rho,x_1)\partial_\mu\varphi\, \beta^{(k)} \cdot T \makebox[.5in]{,}\phi_\alpha=h_\alpha(\rho,x_1) \, e^{i\beta^{(k)} \cdot \alpha \,\varphi}\;,
  \label{ansatz-A}
	\end{align}
where $a, h$ are real profiles.  In order to have a finite action and well-defined regular fields, the appropriate conditions are
$	a(\rho ,x_1)\to 1$, $h_\alpha(\rho , x_1) \to v $, when $\rho \to \infty$. In addition $a(0,x_1)$ and $h(0,x_1)$ must vanish between the quarks and
tend to $1$ and $v$, respectively, on the lines belonging to $S(\mathcal{C}_{\rm e })$, so as to cancel the effect of $J'_{\mu \nu}$. For large separations between the quark-antiquark pair, the saddle-point equations on the plane $x_1=0$ are
\begin{align}
	&\partial_\rho^2 h_\alpha+\frac{\partial_\rho h_\alpha}{\rho} -\frac{(a-1)^2}{\rho^2}(\beta^{(k)}\cdot\alpha)^2=m^2h_\alpha+2\tilde{\lambda} h_\alpha^3\;,\nonumber\\&
	\gamma \vartheta^2\beta^{(k)}\Big(\partial_\rho^2 a+\frac{\partial_\rho a}{\rho}\Big) =2\sum_\alpha \alpha h_\alpha^2 \beta^{(k)}\cdot\alpha(a-1)\;.\label{scalareqs}
\end{align}
The positive roots of $SU(N)$ can be written as $\alpha_{ij}=\omega_i-\omega_j$, with $i<j$. Moreover, the defining weights  have the properties:
\begin{align}
    \omega\cdot \omega = \frac{N-1}{2N^2}\makebox[.5in]{,}
\omega_i\cdot \omega_j = -\frac{1}{2N^2}\makebox[.5in]{,}i\neq j \;.
\end{align}
Then, from Eqs. \eqref{betak-a}, \eqref{betak-b}, we see that the product $\beta^{(k)}\cdot \alpha$ can only take the values $+1$ or $0$. 
 Hence, it is natural to propose the solution 
 \cite{prospecting-paper}:
\begin{eqnarray}
h_\alpha=h \makebox[.5in]{,} \alpha\cdot \beta^{(k)} = 1\;,\nonumber\\
h_\alpha=v \makebox[.5in]{,} \alpha\cdot \beta^{(k)} = 0\;. 
\label{h-free}
\end{eqnarray}
Indeed, because of the property in Eq. \eqref{alphasum}, the ansatz closes. 
Therefore, there is a collective behavior such that only the profiles $h_\alpha$ with $\alpha\cdot \beta^{(k)}=1$ satisfy nontrivial equations of motion. Let us now count these. The condition $\alpha_{ij}\cdot \beta^{(k)}=1$ holds if and only if $i\leq k$ and $j>k$. This results in $k(N-k)$ possibilities and a string tension with  the asymptotic Casimir law
\begin{align}
    \sigma_k = \frac{k(N-k)}{N-1} \sigma_1\;,
\end{align}
where 
\begin{align}
    \sigma_1= \int d^2x \left(\frac{B^2}{2}+D_\mu\bar{\phi} D_\mu \phi +m^2|\phi|^2 + \tilde{\lambda}|\phi|^4\right) 
\end{align}  
is the energy per unit length associated to the Nielsen-Olesen string. Of course, this counting also implies that the tension at the BPS point in Eq. \eqref{tension-bps} also satisfies the Casimir scaling. 
 
 It is important to underline that the Abelian action $S(J')$ is not physically equivalent to the model defined by $W_{\rm mix}[V, \phi]$, which only depends on the value of $k$ (modulo $N$).  In our context, $S(J')$ only applies to a particular solution of the lattice model in the continuum. For the collimated fluxes in center-vortex/monopole chains, the Wilson loop only sees linking numbers. In this case, the probability for a center-vortex branch to leave a monopole in a given direction is distributed isotropically. In contrast, in monopole-only scenarios, it is the flux from the monopoles that spreads isotropically, thus giving 
weight-dependent fluxes through $\mathcal{C}_{\rm e}$.  In this respect, it is instructive to discuss the double-winding Wilson loop for quarks in the defining representation of $SU(N)$.  This observable was proposed in Ref. \cite{d-wl} to differentiate between center-vortex and monopole-only dual superconductor scenarios, as in $SU(2)$ they predict a difference-of-areas and sum-of-areas law, respectively. 
In that work, the full Monte Carlo lattice simulation of $SU(2)$ Yang-Mills theory ultimately established the validity of the first option.

\subsection{Double-winding Wilson loops}
\label{sec-doublewinding}
\begin{figure}[t]
    \centering
    \begin{minipage}{0.5\textwidth} 
        \centering
\includegraphics[width=\textwidth]{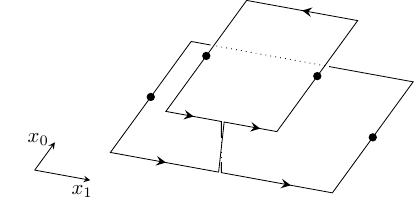} 
        \caption{ Double-winding Wilson loop: the transverse shift is given by $\delta z$. }
        \label{d-wilson}
    \end{minipage}
\end{figure}

\begin{figure}[htbp]
    \centering
    \begin{minipage}{0.45\textwidth}
        \centering
        \includegraphics[width=\textwidth]{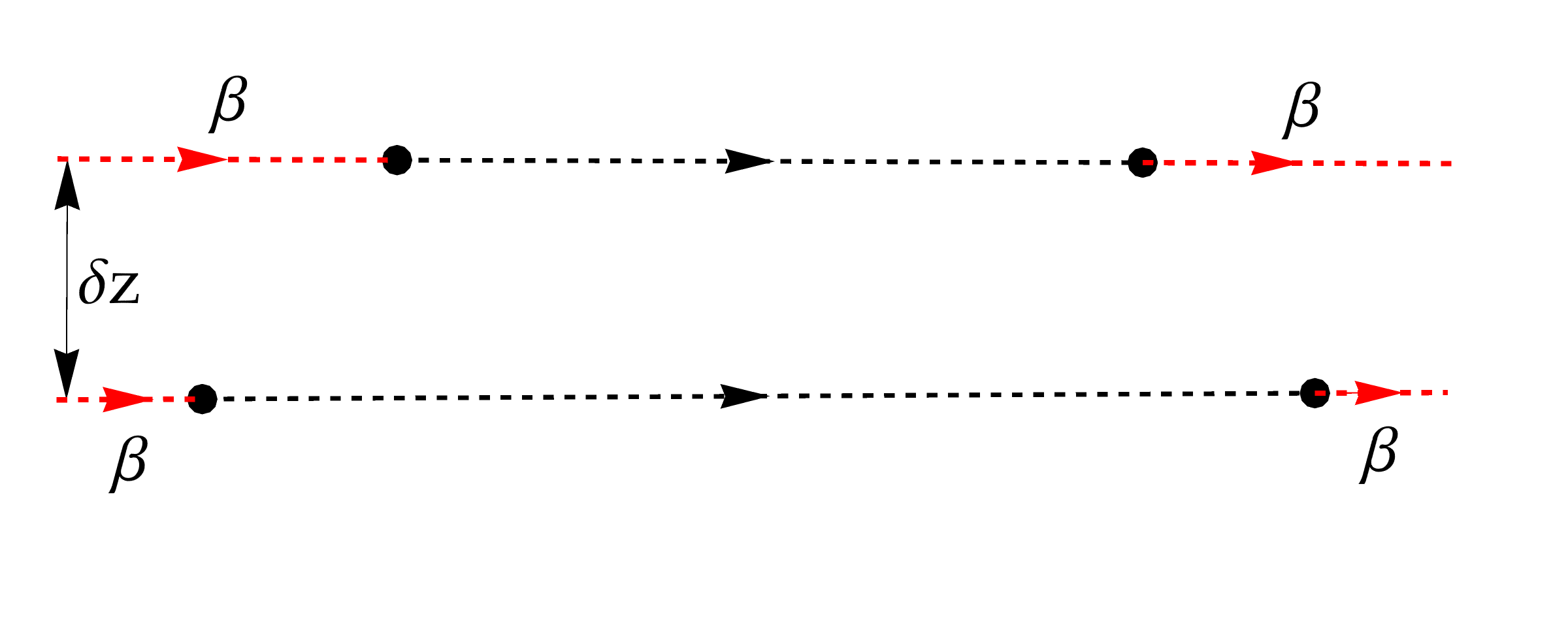} 
        \subcaption{Monopole-only dual superconductor.} 
        \label{fig:figure1a}
    \end{minipage}


    \begin{minipage}{0.48\textwidth}
        \centering
        \includegraphics[width=\textwidth]{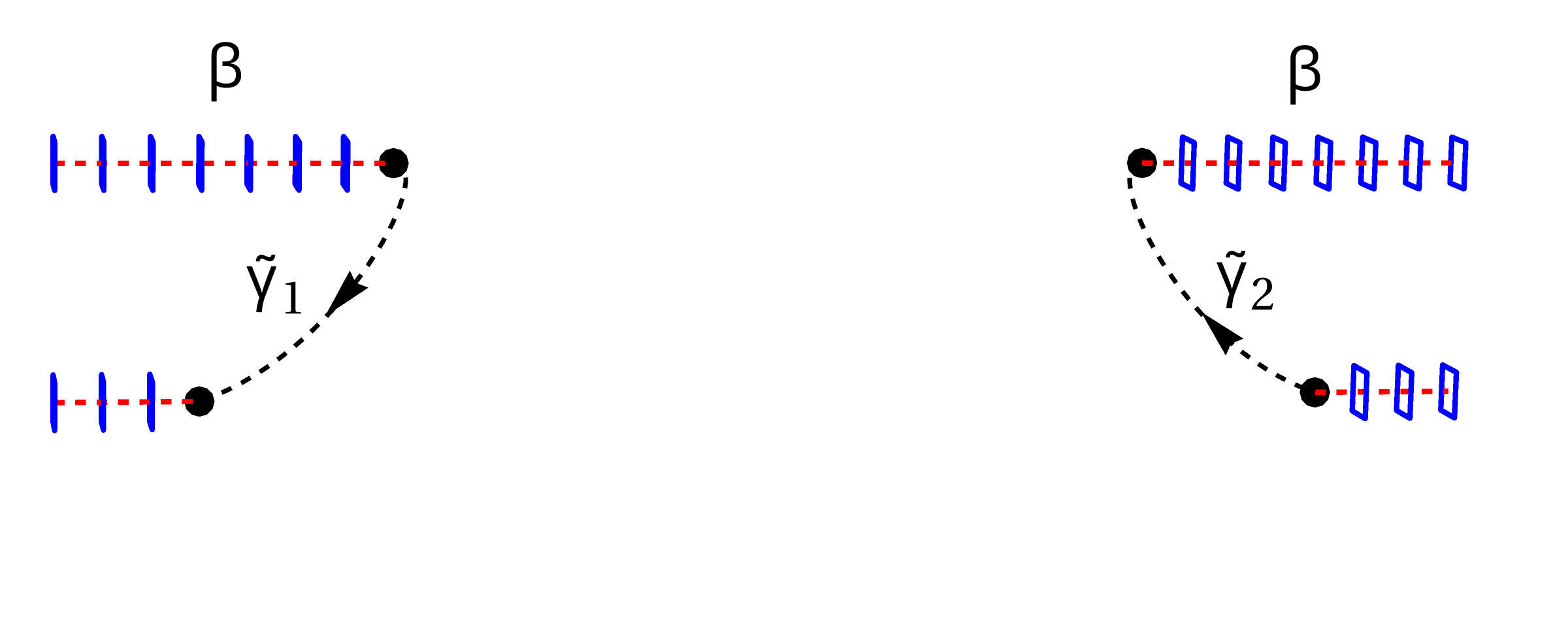} 
        \subcaption{Center vortices and chains: $SU(2)$.} 
        \label{fig:figure2b}
    \end{minipage}
    \hfill
    \begin{minipage}{0.48\textwidth}
        \centering
        \includegraphics[width=\textwidth]{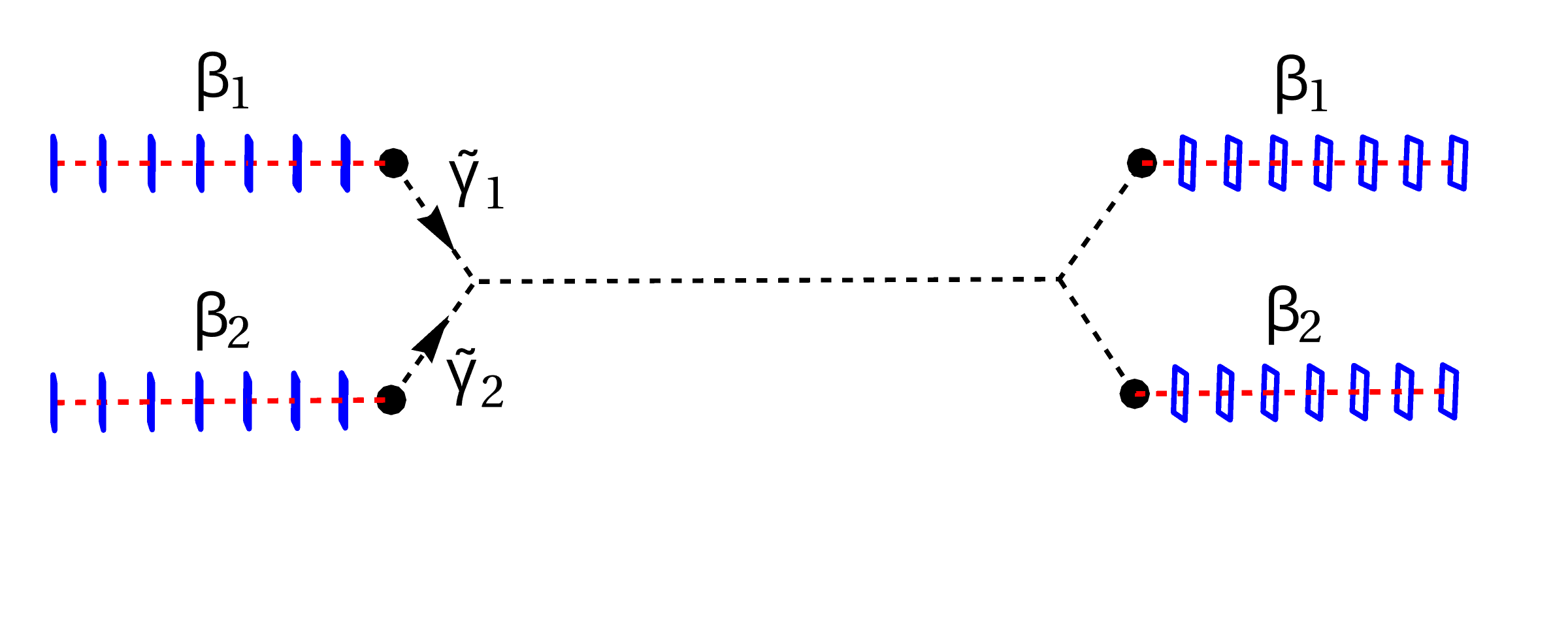} 
        \subcaption{Center vortices and chains: $SU(3)$.  } 
        \label{fig:figure3c}
    \end{minipage}
    \caption{Double-winding Wilson loop at an intermediate time-slice $x_0=0$, where we see two quarks (antiquarks) on the left (right). We use the same conventions of Fig. \ref{fig:wloop}. }
    \label{fig:combined}
\end{figure}

 Let us discuss the  shifted
rectangular double-winding Wilson loop (see Fig. \ref{d-wilson}) in light of the lattice model for the mixed ensemble of center vortices and chains, comparing it with the monopole-only ensemble in Appendix \ref{mon-only}. The latter describes compact Cartan gauge fields $A_\mu$ and a gas of monopole worldlines carrying all possible adjoint charges $\alpha$. In this case,  the Wilson loop for fundamental quarks can be obtained from its dual version, where a gauge field $\lambda_\mu$ is originated after performing a Hubbard-Stratonovich transformation and solving the constraint imposed by the integration over all possible Cartan gauge fields $A_\mu$. At the end, the monopole-only dual action
has the same form as $S(J)$ in Eq. \eqref{S-abe}, with $J_{\mu \nu} =  2\pi\beta\cdot T s_{\mu\nu}$ and $\beta$ being any magnetic weight of the defining representation.   This type of dual superconductor model was introduced in Ref. \cite{maedan} and reviewed in \cite{yanddelta}. For general $SU(N)$, at a fixed intermediate time, as the charge $\beta$ is carried all along the double-winding loop $\mathcal{C}_{\rm e}$,  the sources depicted in Fig. \ref{fig:figure1a} are obtained. As discussed in Ref. \cite{d-wl}, a pair of flux tubes carrying flux
 $\beta$, from left to right, will be induced. When $\delta z$ is lowered, there could be an interaction between them, which depends on the dual superconductor type. Anyway, the associated law will be approximately given by the sum of areas spanned by the flux tubes.   
For the mixed ensemble of center vortices and chains, the lattice action is given by $W_{\rm mix}[V, \phi]$ in Eq. \eqref{Wmix}, with $b(p) = (2\pi/N)\,  s(p)$ and $s(p)$ nontrivial only when $p$ intersects a surface whose border is $\mathcal{C}_{\rm e}$.    Again, when it comes to determining the lowest action, we must determine the best weight distribution for a fixed frustration $e^{ib(p)}$. When $\delta z$ is small, as the continuum is approached, the lowest action will  be characterized by a pair of angles $\xi_1$, $\xi_2$, multivalued with respect to paths $\tilde{\gamma}_1$, $\tilde{\gamma}_2$. They are shown for $SU(2)$ and $SU(3)$ in Figs. \ref{fig:figure2b} and \ref{fig:figure3c}, respectively. In each case, away from the quarks, the fields will be given by Eq. \eqref{delta-con}, with 
\begin{gather}
 \Delta(l) = \left\{ \begin{array}{lll}
 \beta \Delta_l \xi_1 + \beta \Delta_l \xi_2 \;,  & \beta = \sqrt{2}   \\
 \beta_1 \Delta_l \xi_1 + \beta_2 \Delta_l \xi_2 \;, & \beta_1 = (\sqrt{3}, 1) \makebox[.5in]{,} \beta_2 = (-\sqrt{3}, 1 )  \;.
\end{array} \right.  
\end{gather}
Indeed, these configurations give the same frustration in the upper and lower plaquettes, as required: for $SU(2)$,  $ e^{i\pm 2\pi \beta\cdot T}=e^{ \pm i\pi \sigma_3}=-I$, while for $SU(3)$, $e^{i2\pi\beta_1\cdot T}=e^{i2\pi\beta_2\cdot T}=e^{-i\frac{2\pi}{3}}I$. In the first case, the action will be concentrated on a pair of flux tubes on top of the curves $\tilde{\gamma}_1,\tilde{\gamma}_2$ formed by the red and black lines in Fig. \ref{fig:figure2b}, thus leading to a difference-of-areas law. In the second case, as $\beta_1+\beta_2=-\beta_3$, it is clear that, in the global minimization, it is energetically favorable for the flux tubes to coalesce into a single one, leading to the double-Y shaped flux tube depicted in Fig. \ref{fig:figure3c}. In this case, the paths $\tilde{\gamma}_1$ and $\tilde{\gamma}_2$ join at the left junction and continue up to the right junction where they separate. Similarly, the property $\beta_1+\beta_2 = -\beta_3$ implies that the flux tube associated with a configuration of three quark sources in $SU(3)$ forms a Y-shape, in agreement with lattice simulations with Abelian-projected variables \cite{ab-proj-4}.

\subsection{The effect of monopole matching}
\label{three-line}

It is easy to see that the matching of three monopole worldlines at a point is generated by adding the following cubic term in the potential $\mathcal{U}(\phi)$ of Eq. \eqref{Wmix}:
\begin{gather} 
 a c_{\alpha\gamma \delta}\phi_\alpha\phi_\gamma \phi_{\delta}  \;,
 \label{cubic-a}
\end{gather}
where we adopted a summation over repeated indices that considers not only positive but also negative roots, with the condition $\phi_{-\alpha} = \bar{\phi}_{\alpha}$. The coefficients $c_{\alpha \gamma \delta }$ must vanish if the constraint in  Eq. \eqref{m3} is not satisfied. They must also be invariant under root transformations induced by 
permutations of the defining weights $\omega_i$.\footnote{The different roots $\alpha_{ij}$ are given by the $N(N-1)$ possible values of $\omega_i -\omega_j$.} New quartic monopole interactions that contain   $|\phi_\alpha|^2  |\phi_\beta|^2 $ may also be included. 
Note that the ansatz in Eq. \eqref{ansatz-A} continues to close the flux tube field equations
\begin{gather}
D^2\phi_\alpha=m^2 \phi_\alpha + 2\tilde{\lambda}\phi_\alpha(\bar{\phi}_\alpha \phi_\alpha) + 
3  c_{-\alpha \gamma \delta }\, \phi_\gamma \phi_\delta  + \dots \nonumber \;,
\end{gather} 
as the monopole-matching constraint yields
\[
 \phi_\gamma \phi_\delta =  h_\gamma h_\delta\, e^{i \beta \cdot \gamma \varphi} e^{i \beta \cdot \delta \varphi} = h_\gamma h_\delta\, e^{i \beta \cdot \alpha \varphi}  \;. 
\]
As a consequence, even in the Abelian-projected setting, the field profiles are not necessarily of the Nielsen-Olesen type and the asymptotic Casimir scaling could only be approximately satisfied.

\section{Ensembles with local d.o.f.}
\label{nae}

In the previous section, the Abelian-projected scenario was translated from the continuum Hamiltonian formalism to the lattice 4D partition function. The $N$ types of worldsurfaces with global charges were generated
by a complex diagonal $N\times N $ matrix $V(x,y)$. Here, using the Weingarten representation of surfaces and loops, we revisit the 4D  mixed ensemble with non-Abelian d.o.f. of Ref. \cite{mixed} (cf. Eq. \eqref{mixZ}). Our main focus will be
oriented to compare this possibility with the Abelian-projected scenario. Let us initially consider center-vortex worldsurfaces generated by general link-variables $V(x,y) \in \mathbb{C}^{N \times N}$ governed by the lattice action \eqref{wN-m}. As already discussed in Sec. \ref{wsur},  the quartic model generates surfaces with $N$ possible labels at each vertex,  stabilized by contact interactions. In the $\gamma$ expansion, if a given link is only occupied by $N$ variables $V(x,y)$, the contribution is zero. The situation changes when this is combined with a factor originated from the expansion of a new term in the action $\xi (\det V(x,y)+\det V(y,x))$. Because of this $N$-matching property, the symmetry is reduced to 
\begin{align}
	V(x,y)\to U(x) V(x,y) U^{-1}(y) \makebox[.5in]{,} U \in SU(N)\;.
	\label{n-a-symmetry}
\end{align}
To generate chains, we can proceed as in Sec. \ref{abe-chains}, but taking into account the non-Abelian properties of $V(x,y)$. The monopole loops are again introduced according to the discussion in Sec. \ref{eloops}. They are generated by monopole fields $\zeta_\alpha $
in the adjoint representation of $SU(N)$
\begin{gather}
    Z_{\rm mix}[b] \propto \int DV D\zeta \,\exp\left(-W_{\rm mix}[V,\zeta]\right) \makebox[.5in]{,} 
    W_{\rm mix}[V,\zeta]=W_{\rm c.v.}[V]-\sum_l \tilde{V}(l)+\tilde{Q}[\zeta]\nonumber\\
    \tilde{V}(l)= \sum_\alpha \zeta^\dagger_\alpha (x)
    H (x,y)\,  \zeta_\alpha (y)
    \makebox[.5in]{,}   H(x,y)|_{AB}= {\rm Tr}(V(x,y)T_B V^\dagger(x,y) T_A) \;, \nonumber \\ \tilde{Q}[\zeta] =   \sum_x \sum_\alpha \tilde{\mathcal{Q}}(\zeta_\alpha(x)) 
\makebox[.5in]{,}  \tilde{\mathcal{Q}}(\zeta)  =   \tilde{\eta} \,  \zeta^\dagger \zeta +
\tilde{\lambda}\,  (\zeta^\dagger \zeta)^2\;.
\label{mix-sem}
\end{gather} 
The interpolation property between pairs of center-vortex worldsurfaces can be analyzed as already done for the Abelian-projected case. This time,  expanding in $\tilde{V}(l)$ and integrating over $V$, there are contributions that involve the loop variables
\begin{align}
{\rm Tr}\,   H(x_1,x_2)\dots H(x_{n},x_1)\;. 
\end{align}  
In addition,  Eq. \eqref{n-a-symmetry} implies
\begin{gather}
	H(x,y)\to  
	 R(x)   H(x,y) R^{-1}(y) 
	\makebox[.5in]{,} U T_A U^{-1} = T_B R(U)|_{BA} \;,
\end{gather}
so that the model still has the  local symmetry \eqref{n-a-symmetry}, as long as 
\begin{gather}
\zeta_\alpha (x)\to  
R(x) \zeta_\alpha (x) 
\end{gather}
($R = R(U)$ is the adjoint representation of $U$). 
Therefore, the ensemble $Z_{\rm mix}[b]$ also contains center-vortex worldsurfaces attached to monopole worldlines, which are associated to 
$SU(N)$-singlets in the
$\gamma$ expansion. Now, we can follow the same series of steps as in the Abelian-projected case. As discussed in the general Weingarten framework, when (center-vortex) worldsurfaces percolate, $V(x,y)$ is forced to be close to $\vartheta U(x,y)$, $U(x,y) \in U(N)$. In addition, we shall assume the effect of $N$-matching to be strong (large $\xi$), which drives $U(x,y)$ close to $SU(N)$. Then, when this initial stage dominates, the monopole sector can be approximated by Eq. \eqref{Rg} with $\tilde{\gamma}=\vartheta^2$ and $R(x,y) = R(U(x,y))$, $U(x,y) \in SU(N)$. That is, the model becomes 
\begin{gather}
W_{\rm mix}[V, \zeta] \approx  \gamma \vartheta^2   \sum_{p}{\rm Tr} \Big( I -  e^{i b(p)}  U(p) \Big) +  \vartheta^2\sum_{x,\mu} \sum_\alpha (\Delta_\mu \zeta_\alpha)^\dagger \Delta_\mu \zeta_\alpha + \mathcal{U}(\zeta)
     \;, \label{gen-m} \\
    \Delta_\mu \zeta =
       R(x, x+ \mu) \zeta(x+ \mu)  - \zeta(x) \makebox[.3in]{,}  \mathcal{U}(\zeta) = \sum_x
  \sum_\alpha  \left(a^2 m^2  |\zeta_\alpha|^2  
    + \tilde{\lambda}\,  |\zeta_\alpha|^4 \right)\;.
       \label{nonam}
\end{gather}

\subsection{Non-Abelian scenario and  the confining flux tube} 
\label{nonaf}

The different adjoint flavors $\zeta_\alpha$ in Eq. \eqref{nonam} are uncorrelated, so that $N(N-1)$ independent ensembles of loops ${\rm Tr}\, \Gamma_R(\mathcal{C})$ labeled by the adjoint weights $\alpha$ are generated (see Eq. \eqref{R-holo}). Up to this point, there is only repulsion between worldlines of the same type $\alpha$. However, this could compete with other interactions and natural correlations, such as the monopole matching properties considered in Ref. \cite{mixed} (see Sec. \ref{m-match}). Cubic and quartic interactions were exhaustively reviewed and analyzed in Ref. \cite{prospecting-paper}.  In fact, in the non-Abelian case,  they are essential to drive a phase that supports the formation of a confining flux tube. Otherwise, the vacua manifold  would be the product of $N(N-1)/2$ higher dimensional spheres $\bar{\zeta}_\alpha^A \zeta_\alpha^A = v^2$ (no summation over $\alpha$). In the continuum limit this would imply a trivial first homotopy group. In the non-Abelian lattice scenario, three-line matching may be included by the cubic term:
\begin{gather}
   a\kappa N_{\alpha \gamma } (\zeta^\dagger_{\alpha + \gamma}, [\zeta_{\alpha}  , \zeta_{\gamma}]) =  a\kappa   N_{\alpha \gamma 
    \delta} (\zeta_{\delta}, [\zeta_{\alpha}  , \zeta_{\gamma}]) \makebox[.5in]{,} \zeta_{-\alpha}= \zeta_\alpha^\dagger \;.
    \label{cubic-no}
\end{gather}
Here, instead of a column matrix with components $\zeta^A$, we changed the notation to use  Lie algebra valued fields $\zeta = \zeta^A T_A$. The structure constants follow the conventions of Ref. \cite{prospecting-paper}. As usual, $N_{\alpha \gamma } $ is only nonzero when $\alpha + \gamma $ is a root. This root can always be written as $-\delta$, where $\delta$ is a root. This defines the structure constants in the second member that only contribute when $\alpha + \gamma + \delta =0$.  In this notation, the covariant derivative and the transformation of monopole fields read
\begin{align}
	 \Delta_\mu \zeta =
U(x, x+ \mu) \zeta(x+ \mu) U(x+ \mu ,x) - \zeta(x) \makebox[.5in]{,} \zeta (x)\to U(x)\zeta(x) U^{-1}(x)\;.
\end{align} 
The important point is that the additional interactions can easily generate a monopole condensate such that the system  undergoes $SU(N) \to Z(N)$ SSB \cite{prospecting-paper}, which gives a vacua manifold with nontrivial first homotopy group 
 \begin{gather}
     \Pi_1(SU(N)/Z(N)) = Z(N) \;.
     \label{Pi1}
 \end{gather} 
In the continuum limit, this will lead to topological solutions with $N$-ality properties, such as confining flux tubes with center-charges ($k$-strings)  \cite{David,it,shif,David2,marshakov,PhysRevD.71.045010}.
In the lattice,  similarly to the Abelian-projected case, the following conditions must be approximately satisfied
away from the quarks:
\begin{gather}
	U(p) = e^{-ib(p)} I \makebox[.5in]{,}   U(x,y)\zeta_\alpha(y) U(y,x)=\zeta_\alpha(x)    \;.\label{bc-nonabe}
\end{gather}
In addition, the monopole fields must be close to the vacua manifold, which can be assumed to be $\zeta_\alpha = v U E_\alpha U^{-1}$. Although this is not the most general form, it occurs in a large class of  $SU(N) \to Z(N)$ SSB models. By the same arguments presented in the Abelian case, we must have 
\begin{gather}
 U(x,y) = U(x) \, e^{i \Delta(l) \cdot T } U^{-1}(y) \makebox[.5in]{,}  \Delta(l) = \beta' \Delta_l \xi   \;,
\end{gather}
where $\Delta_l \xi$  was defined after Eq.  \eqref{Delta-defi} and $U(x) \in SU(N)$ is an arbitrary field defined at the sites. That is, a general weight $\beta'$ with $N$-ality $k$ comes into play, not necessarily equal to $\beta_{\rm e}$, to be determined by lowering the associated saddle-point action. 
Again, the monopole fields are consistently defined at the sites as starting from $\zeta_\alpha(x)$, applying Eq. \eqref{bc-nonabe} to consecutive links along a closed path,  we return to $\zeta_\alpha(x)$. When $k=0$, there is no frustration and no confining string. 
For nontrivial $k$, with $ U_\mu(x,y) = e^{ia \Lambda_\mu(x)}$, $ \Lambda_\mu \in \mathfrak{su}(N) $, the continuum version of the saddle-point equations can be used to determine which $\beta'$ leads to the lowest action
\begin{align}
 &S(J')=\int d^4x \left( \frac{\gamma\vartheta^2}{4} ( F_{\mu\nu}- J'_{\mu\nu})^2  + \frac{1}{2} \Big( (D_\mu \zeta_\alpha)^\dagger , D_\mu \zeta_\alpha \Big) + \mathcal{U}(\zeta) \right)  \;,
 \label{Sjota}
\end{align} 
where $J'_{\mu \nu}$ was defined in Eq. \eqref{S-abe-a}.
These equations were extensively analyzed in  Ref. \cite{prospecting-paper}, with the lowest action 
given by the antisymmetric weight 
$\beta^{(k)}$. Similarly to the 
Abelian-projected case, as the parameters are changed, these models allow for different 
flux tube profiles that could fit the lattice data produced at asymptotic distances.

\subsection{Comparing Abelian-projected and non-Abelian scenarios}
\label{comparing}
To better compare the scenarios, note that the Abelian-projected model 
\eqref{abz}
is embedded in the non-Abelian one by simply considering $V(x,y)$ diagonal and $\zeta_\alpha = \phi_\alpha E_\alpha$. In this case\footnote{The only nontrivial element of the matrix $E_\alpha$, $\alpha_{ij} =\omega_i - \omega_j$, is at the $ij$ position.}
	\begin{align}
		\tilde{V}(l) =  \sum_\alpha  \bar{\phi}_{\alpha}(x) \phi_{\alpha}(y) {\rm Tr}(V(x,y) E_{\alpha} V^\dagger(x,y) E^\dagger_{\alpha}) = \frac{1}{2N}\sum_{i<j}\bar{\phi}_{ij}(x) \phi_{ij}(y) \bar{V}_j(x,y)V_i(x,y) \;. 
	\end{align} 
	In particular, when center vortices condense, Eq. \eqref{Wmix} can be encoded in Eqs. \eqref{gen-m}, \eqref{nonam} with
	\begin{gather}
		U(x,y) = \left(
		\begin{array}{cccc}
			U_1 & 0 & \cdots & 0 \\
			0 & U_2 & \cdots & 0 \\
			\vdots & \vdots & \ddots & \vdots \\
			0 & 0 & \cdots & U_N
		\end{array}
		\right) \makebox[.5in]{,} \prod_i U_i =1 \makebox[.5in]{,} \zeta_\alpha = \phi_\alpha E_\alpha  
	\end{gather}
(no summation over $\alpha$). Also, the cubic term 
\eqref{cubic-a} is embedded in the form \eqref{cubic-no}, with 
$c_{\alpha \gamma \delta} =  \kappa N^2_{\alpha \gamma \delta}$. When $c_{\alpha \gamma \delta }$ is negative, the fusion of three monopole charges is favored. This corresponds to
$\kappa < 0$, which is also required to drive the above mentioned $SU(N) \to Z(N)$ SSB pattern. Now, let us consider the mixed ensemble partition function \eqref{gen-m}, with 
$\mathcal{U}(\zeta)$ containing  the additional interactions required to drive this pattern. Of course, all possible configurations, Abelian-like or not, will be present. However, when computing the Wilson loop, the relevant ones are expected to be distributed around 
a flux tube characterized by the antisymmetric weight with the same $N$-ality of the quarks representation $D(\cdot)$. Indeed, as shown in Ref. \cite{prospecting-paper}, there is a region in parameter space where the solutions have the Abelianized form $ \zeta_ \alpha = \phi_\alpha E_\alpha $, together with Eq. \eqref{ansatz-A}. This is a stable region in the sense that, under small deviations, the action continues to be bounded from below and the associated properties are only perturbatively modified. That is, for the Wilson loop calculation, the relevant center-vortex worldsurfaces with $N$ local ``colors'' at the vertices essentially behave as $N$ surface types with global charges $\beta_i$.  For this reason, in the lattice, the Abelian-projected and non-Abelian models share various properties at asymptotically large distances between quark probes.

\subsubsection{Common features}
\label{common}

Here, we list the main physical properties of the Abelian-projected scenario obtained in Sec. \ref{ape}, reference previous works where these properties have been discussed in the non-Abelian context, and briefly comment on the available lattice evidence. 

\begin{itemize}

\item In Sec. \ref{Abe-ft}, in the simplest Abelian-projected ensemble where monopoles of different species do not interact, we showed that:

- $N$-ality is  manifested in the form of an asymptotic Casimir law, $\propto k(N-k)$. In the non-Abelian scenario, this law was obtained for a particular model in Ref. \cite{oxgustavo} and showed to be stable at the BPS point in Sec. V A of Ref. \cite{O-S-J}. 
It was also obtained in models where the Higgs-field content is only given by the complex adjoint fields $\zeta_\alpha$ in Eq. \eqref{Sjota}, see Sec. VI B of Ref. \cite{prospecting-paper}. Present lattice data place the string tension somewhere between this behavior and the sine law \cite{lucini-teper,latt-scaling}. 
    
   - The transverse profiles are  $k$-independent and display a Nielsen-Olesen transverse distribution. The non-Abelian counterpart was initially discussed in Refs. \cite{oxvercauteren,oxgustavo} and then generalized to a large class of effective models in Sec. V of Ref. \cite{prospecting-paper}. As previously discussed, after disregarding the effect of center vortex thickness on the Wilson loop, these models are expected to be applicable at asymptotic distances. However, the distances considered in the analysis of the transverse distribution of the 4D  $SU(3)$ YM energy-momentum tensor and field profiles are intermediate to nearly asymptotic \cite{kitazawa}. The lattice data show deviations from the Nielsen-Olesen behavior, which tend to decrease at larger distances. Additionally, in Ref. \cite{Cosmai-2017}, the longitudinal chromoelectric field of the flux tube was fitted with a Nielsen-Olesen profile. However, lattice simulations exhibit larger error bars in the asymptotic regime, so there is no conclusive evidence yet regarding the field profiles in this region.

   \vspace{.05cm}

\item In Sec. \ref{sec-doublewinding}, we discussed other manifestations of $N$-ality. Namely, the difference-of-areas law for double Wilson loops in $SU(2)$  and double Y-shaped potentials in $SU(3)$. For the non-Abelian case, they were discussed in Sec. VII of Ref. \cite{mixed} and Sec. V of  Ref. \cite{O-S-J}. In the lattice, these properties were observed  in Refs.  \cite{d-wl} and \cite{doubleY}, respectively.

\item In Sec. \ref{three-line}, we showed that, in the Abelian setting, departures from the exact Casimir law and Nielsen-Olesen profiles can be parametrized as due to three-line monopole matching.  This can also change the phase transition to be first order. This type of matching only exists for $N \geq 3$. For $N=2$ there are just two (one-component) roots $\alpha = \pm 1/\sqrt{2}$. It is satisfying to note
    that the confinement/deconfinement phase transition is known to be first (second) order for $N \geq 3$ ($N=2$). Deviations from the Casimir law have been observed in Ref. \cite{latt-scaling}.   
    
\end{itemize}

\subsubsection{Differences}
\label{diff}

In the continuum, unlike the Abelian-projected case, where $S(J')$ in Eq. \eqref{S-abe} is only used to evaluate the action of a particular solution,  in the non-Abelian case it can also be used for the global minimization problem. This is because of the topological properties implied by Eq. \eqref{Pi1}. Even if $\beta' $ is not an antisymmetric weight $\beta^{(k)}$,  the global minimization will be able to transition to $\beta^{(k)}$. For example, for adjoint quarks, if we start with $\beta_{\rm e} = 2N \alpha$, the minimization gives a 
vanishing action. This is attained for a well-defined  non-Abelian single-valued solution 
\begin{gather}
 \Lambda_\mu = iS \partial_\mu S^{-1} \makebox[.5in]{,}  \zeta_\alpha = v S E_\alpha S^{-1}  \;,
\end{gather}
which cancels the effect of the sources (see Ref. \cite{proc-oxman}). Therefore, in the non-Abelian case, the continuum limit of $Z_{\rm mix}[b]$ is expected to be captured by the partition function:
\begin{eqnarray}
	Z_{\rm mix} [J_{\rm e}] =  \int [{\cal D}\Lambda_\mu] [{\cal D }\zeta ] \,   e^{-S(J_{\rm e}) } \makebox[.5in]{,} J_{\rm e}|_{\mu \nu} = 2\pi\beta_{\rm e}\cdot T s_{\mu\nu}\;.
\end{eqnarray} 
Another difference relates to the possibility of extending the model to the intermediate confining region, where the flux tube 
tension scales with the quadratic Casimir of the quark representation \cite{casimir-bali}. In this regime, the total energy stored in the adjoint string is smaller that the energy required to screen the adjoint quark probes \cite{adjoint-breaking}. It is well-known that Abelian projection cannot describe this regime \cite{noabelianadjointtube}. For a general ensemble formed by thick configurations,  when considering the adjoint
representation, not all the terms in Eq. \eqref{wloope} give the same average. In this case, there are always vanishing weights, so that the average $\langle {\rm Ad}  (W_{\mathcal C})  \rangle$ is never exponentially suppressed. On the other hand,  the combined effect of center-vortex stiffness and domains characterized by general non-Abelian orientations was advocated as a possible explanation for the intermediate Casimir scaling \cite{casimirscalingthickness,casimirscalingthickness-2}. Then, in the non-Abelian ensemble, there is still the possibility to include the effect of center-vortex thickness to get a smoother transition from the asymptotic to the intermediate region. An investigation in this direction will be presented elsewhere.

\section{Conclusions}
\label{conclusions}

In this work, we initially derived an effective lattice representation for the partition function associated with an Abelian-projected ensemble of center vortices and chains, complementing the previously obtained wavefunctional representation. These objects and some of their properties have already been detected in $SU(2)$ and $SU(3)$ lattice studies. On this direction, further lattice investigations will be valuable to refine the ensemble characterization. Here, a key conclusion is that these objects could dominate the Abelian-projected lattice configurations. On the one hand, it is known that Abelian-projected variables obtained from the full Monte Carlo configurations capture $N$-ality at asymptotic distances \cite{ab-proj-1,ab-proj-2}, the flux tube for fundamental quarks  in $SU(2)$ and $SU(3)$ \cite{ab-proj-3,suga2014},  and the three-quark potential in $SU(3)$ \cite{ab-proj-4}. From our modeling, all these properties are also successfully described, showing how an Abelian ensemble of topological objects is compatible with $N$-ality while simultaneously supporting a ``dual superconductor'' description of the fundamental string. The origin of the gauge-field in this description is different from the usual one obtained in monopole-only scenarios, which is generated after a Hubbard-Stratonovich transformation and  solving the constraint implied by integration over all Abelian gauge fields (see Appendix \ref{mon-only}). In our modeling, the lattice gauge field $U_\mu \in U(1)^{N-1}$ emerges associated to Goldstone modes in an ensemble of percolating center-vortex worldsurfaces. This was evidenced by the Weingarten matrix representation of surfaces, which also helped to elucidate the non-Abelian Goldstone modes when the surfaces are also characterized by local labels.  In this respect,  we initially noted that the representation based on $V(x,y) \in \mathbb{C}^{N\times N}$ matrix link-variables with Gaussian weight describes the ensemble of closed noninteracting surfaces with $N$ possible labels at their vertices.  This noninteracting model served as a basis to include the properties that turn the surfaces into the guiding centers of center vortices and chains. As is well-known, the Gaussian model is in fact unstable. Then, the next step was to introduce an appropriate stabilizing quartic term that provides contact interactions. 
This way, not only the normal but also the percolating phase was stabilized. Here, the interesting result is that, as the effect of the quartic term gets stronger, the percolating phase becomes described by gauge field 
link-variables governed by the Wilson action with $U(N)$ local symmetry. In this manner, we extended
 the result of Ref. \cite{Rey} to the non-Abelian case. 
 
  In the center-vortex context, the elementary  objects are characterized by magnetic weights $\beta_i$ of the defining representation, which are physically equivalent.  In the Abelian-projected case, these weights are globally defined on a given center-vortex branch, so the
 ensemble was generated by $N$ variables $V_i(x,y) \in \mathbb{C}$ with a symmetry under the permutations of labels (Weyl symmetry). The proposed ensemble essentially extends the center-vortex lines and pointlike monopoles, as described in the wavefunctional approach, to the 4D partition function formalism, where they become worldsurfaces and worldlines, respectively. One main defining property of center vortices, namely, the generation of a center element when the Wilson loop $\mathcal{C}_{\rm e}$ is linked,  corresponds to an appropriate lattice frustration, which  only depends on the $N$-ality of the quark representation. This relates to the assumption that the Wilson loop is sufficiently large, such that the dominant contribution arises from the linking numbers between the center-vortex guiding centers and  $\mathcal{C}_{\rm e}$. Another fingerprint is the natural matching between $N$ different center-vortex fluxes $\beta_i$,  due to the  property $\beta_1 + \dots + \beta_N =0$. This process was incorporated within the Weingarten representation by means of a term  $\propto V_1(x,y) \dots V_N(x,y)$, which reduces the local symmetry from $U(1)^N$ to  $U(1)^{N-1}$. Similarly to Ref. \cite{mixed}, center-vortex worldsurfaces with different weights $\beta_i$, $\beta_j$ attached in pairs to monopole worldlines (chains) were simply introduced through additional scalar fields. When $N$-matching is strongly favored, the percolating phase can be approximated by the Wilson action for  Goldstone gauge fields $U_\mu \in U(1)^{N-1}$ minimally coupled to the monopole scalars.  Therefore,  the elementary center vortices and chains in 4D provide emergent gauge fields, scalars, and minimal coupling. As the continuum is approached, this field-content is fundamental for generating a topologically stable flux tube. This occurs when not only center vortices percolate, but monopoles also proliferate (condense). The minimal area law with Casimir behavior derived this way is consistent with that obtained in the wavefunctional formalism.  
 This is the simplest lattice model, with fewer parameters, which can describe or approximate the main asymptotic properties of confinement. Natural additional processes, like three-line monopole matching, could introduce deviations from the Casimir law compatible with recent simulations in Ref. \cite{latt-scaling} and even accommodate a first-order phase transition.  This line of thought parallels the one proposed in Ref. \cite{mixed} for an ensemble with non-Abelian degrees of freedom, which we revisited here using the Weingarten representation. In this case, the center-vortex worldsurfaces are generated by a single matrix-valued variable $V(x,y) \in \mathbb{C}^{N\times N}$, where the $N$ possible labels at each vertex correspond to a local  magnetic charge $\beta_i$ that can change from point to point over the surface.   In this case, $N$-matching changes the local symmetry from $U(N)$ to $SU(N)$. In the phase where percolation and $N$-matching dominate, the variables 
 become approximated by $SU(N)$ lattice gauge fields, while chains are generated by minimally coupled
 adjoint monopole fields. 
This time, when monopoles proliferate,  additional natural correlations are essential to generate a flux tube, as already discussed in detail in previous works. Among them, the matching of three monopole worldlines carrying adjoint weights such that $\alpha + \gamma + \delta = 0 $  and additional quartic interactions can drive a monopole condensate with $SU(N) \to Z(N)$ SSB. Because of this pattern, the associated continuum description not only facilitates the computation of the lowest-action configurations but also constitutes a model with $N$-ality properties on its own. Its rich parameter space includes a region where the solutions follow an Abelian-like ansatz with Nielsen-Olesen profiles and asymptotic Casimir law \cite{prospecting-paper}.  Moreover, the $U(1)^{N-1}$ symmetric Weingarten matrix model is naturally embedded in the Cartan sector of the $SU(N)$ symmetric description. In other words, although the non-Abelian model is initially based on local magnetic weights, due to the series of phase transitions, it may undergo a ``dynamical Abelianization'' and behave as having global weights on the center-vortex branches.  Also, in both Abelian and non-Abelian settings, lattice frustration is trivial for adjoint quarks, resulting in a vanishing potential. This aligns with the fact that the models are based solely on averages of center elements, which are trivial in the adjoint representation. In conclusion, the physics described in both settings is quite similar.  Nonetheless, the non-Abelian scenario might offer the potential for implementing some improvements. For example, at asymptotic distances, the potential for adjoint quarks should not be simply zero, but rather a constant representing the energy required to create the pair of valence gluons that screen the quark probes. This corresponds to an appropriate Wilson loop asymptotic perimeter law.
In Refs. \cite{conf-qg,proc-oxman}, this was reproduced by introducing additional adjoint fields, which result in confined adjoint monopole solutions with nontrivial energy that represent the valence gluons.  It would be interesting to further investigate how these fields might emerge at the ensemble level. One natural possibility would be a relation with the off-diagonal $W$-modes, which are also expected to describe this breaking \cite{d-wl} and are erased in Abelian-projected variables. Another possibility would be to establish a smoother connection with confinement at intermediate distances, where Abelian projection is known to be inadequate. Indeed, a potential explanation for intermediate Casimir scaling involves non-Abelian degrees of freedom and the effect of finite center-vortex thickness \cite{casimirscalingthickness,casimirscalingthickness-2}. Finally, as both settings are able to interpolate the percolating and normal phases, it would be interesting to analyze whether they can capture the geometry of center-vortex worldsurfaces observed at finite temperature in Ref. \cite{leinweber-nmatching}.

\newpage 

\appendix

\section{Surfaces in contact} 
\label{doc}

\begin{figure}[t]
		\centering 
		\includegraphics[scale=1]{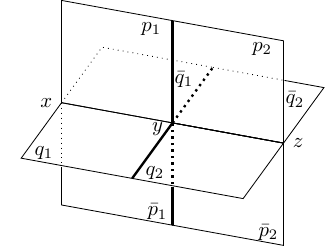}  
	\caption{Two consecutive arrays of four  plaquettes (oriented as in Fig. \ref{fig7:a}).}
	\label{fig9}
\end{figure}

\begin{figure}[t] 
	\centering
	\begin{subfigure}{.5\textwidth}
		\centering
		 \includegraphics[scale=1]{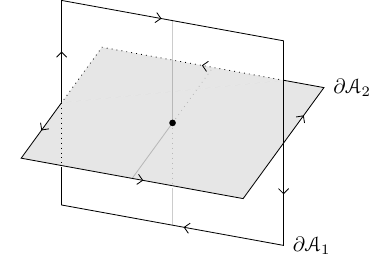}
	\end{subfigure}%
	\begin{subfigure}{.5\textwidth}
		\centering
	 \includegraphics[scale=1]{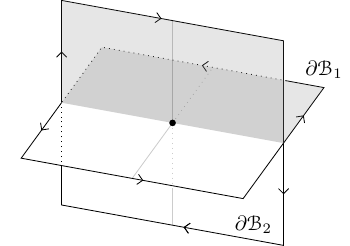}
	\end{subfigure}
	\caption{ The paths $\partial\mathscr{A}_1,\partial\mathscr{A}_2,\partial\mathscr{B}_1,\partial\mathscr{B}_2$.}
	\label{fig8}
\end{figure}

\begin{figure}
	\centering
	\begin{subfigure}{.5\textwidth}
		\centering
		 \includegraphics[scale=1]{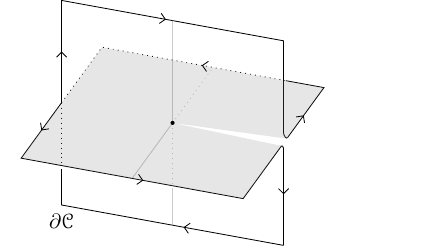}
	\end{subfigure}%
	\begin{subfigure}{.5\textwidth}
		\centering
		 \includegraphics[scale=1]{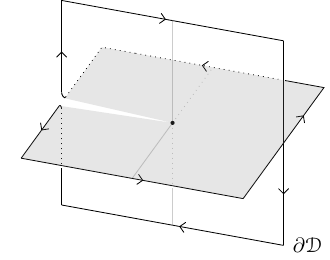}
	\end{subfigure}
	\caption{The paths $\partial\mathscr{C},\partial\mathscr{D}$.  }
	\label{fig10}
\end{figure}

For the array of four plaquettes depicted in Fig. \ref{fig7:a}, using the Gaussian integral  
\begin{align}
	\int dV \,   {\rm Tr} ( V A)   {\rm Tr}( V^\dagger B)  {\rm Tr}( V C) {\rm Tr}( V^\dagger D)\, e^{-\mathcal{Q}_0(V)}  \nonumber \\
	= {\rm Tr} (AB) {\rm Tr} (CD) + {\rm Tr} (AD){\rm Tr} (CB) \;,
	\label{doubleo}
\end{align} 
we get the partial contribution
\begin{align}
	C(p, \bar{p}, q,\bar{q}) = \int dV(x, y)\, {\rm Tr} V(p)  {\rm Tr} V(\bar{p})   {\rm Tr} V(q)  {\rm Tr} V(\bar{q})\, e^{-\mathcal{Q}_0(V(x,y))} \nonumber \\
	= {\rm Tr}(\Gamma(\partial(p \cup \bar{p}))) 
	{\rm Tr}(\Gamma(\partial(q \cup \bar{q}))) +
	{\rm Tr}(\Gamma(\partial(p \cup \bar{q}))) 
	{\rm Tr}(\Gamma(\partial(q \cup \bar{p})))\;.
	\label{a4}
\end{align}
The surface elements associated with the first term (Fig. \ref{fig7:b}) have two edges placed at the same link $(x,y)$, where $p \cup \bar{p}$ intersects $q \cup \bar{q}$. Those associated with the second term (Fig. \ref{fig7:c}) have two edges at  $(x,y)$, where $p \cup \bar{q}$ touches $q \cup \bar{p}$. Now, let us consider two consecutive arrays of four plaquettes (see Fig. \ref{fig9}).
 In the partial contribution, besides the integrals at the doubly occupied links $(x,y)$, $(y,z)$ on $\mathcal{C}$, we also perform the integrals over the matrix variables $V_i$ at the four single  links where these arrays meet (shown in bold). This implements the ``gluings'' $p_1 \cup p_2$, $\bar{p}_1 \cup \bar{p}_2$, $q_1 \cup q_2$, $\bar{q}_1 \cup \bar{q}_2$ and yields
\begin{align}
&\int \prod_{i=1}^4 dV_i \,  C(p_1, \bar{p}_1, q_1,\bar{q}_1) \, C(p_2, \bar{p}_2, q_2,\bar{q}_2)
\nonumber \\
& =  N^2 {\rm Tr} (\Gamma(\partial \mathscr{A}_1)) {\rm Tr} (\Gamma(\partial \mathscr{A}_2)) +   N^2 {\rm Tr} (\Gamma(\partial \mathscr{B}_1)) {\rm Tr} (\Gamma(\partial \mathscr{B}_2))   + N  {\rm Tr} (\Gamma(\partial \mathscr{C}) ) +   N {\rm Tr} (\Gamma(\partial \mathscr{D})) \;.
\label{8array}
\end{align} 
The different paths are shown in Figs. \ref{fig8} and \ref{fig10}, while the surface elements are given by
\begin{gather}
 \mathscr{A}_1 =p_1 \cup \bar{p}_1\cup\bar{p}_2\cup p_2 \makebox[.5in]{,}
\mathscr{A}_2 =q_1\cup \bar{q}_1\cup \bar{q}_2\cup q_2 \nonumber \\
\mathscr{B}_1 =p_1\cup \bar{q}_1\cup \bar{q}_2\cup p_2 \makebox[.5in]{,}
\mathscr{B}_2 =q_1\cup \bar{p}_1\cup \bar{p}_2\cup q_2 \nonumber \\
\mathscr{C} =p_1\cup \bar{p}_1\cup \bar{p}_2\cup q_2\cup q_1\cup \bar{q}_1\cup \bar{q}_2\cup p_2 \nonumber \\
\mathscr{D} =p_1\cup \bar{q}_1\cup \bar{q}_2\cup q_2\cup q_1\cup \bar{p}_1\cup \bar{p}_2\cup p_2  \;.
\end{gather}
Here, the ordering specifies how the plaquettes are glued, which defines different orientations (in these sequences, the last plaquette is also glued to the first). 
Note that at the lattice site $y$ common to the eight plaquettes,
 $\mathscr{A}_1$ and $ \mathscr{A}_2$  ($\mathscr{B}_1$ and $  \mathscr{B}_2$) provide a couple of independent vertices that happen to be at the same place. On the other hand, $\mathscr{C}$ (or $\mathscr{D}$) is a connected single surface element with just one vertex at $y$. Accordingly, in the first case, the contribution in Eq. \eqref{8array} is given by the product of two independent factors like in Eq. \eqref{GA}, while in the second case it is given by just one factor of this type. That is, the factors at $y$ are $N^2$ or $N$, depending on whether the number of vertices at that site is two or one, respectively. In addition, in all the cases, the
number of edges at $\mathcal{C}$ is twice the number of links. 
For example, the surface element $\mathscr{D}$ has a pair of edges at $(x,y)$, where $p_1 \cup \bar{q}_1$ touches $q_1 \cup \bar{p}_1$, and a second pair at $(y,z)$, where $\bar{q}_2 \cup q_2$ intersects $\bar{p}_2 \cup p_2$ (self-intersection). These conclusions can be generalized to the case where $\mathcal{C}$ turns at a right angle or stops growing. In the quartic model, new surface elements are generated, as Eq. \eqref{doubleo} is replaced by 
\begin{align}
	& \frac{1}{z}\int dV \,   {\rm Tr} ( V A)   {\rm Tr}( V^\dagger B)  {\rm Tr}( V C) {\rm Tr}( V^\dagger D) \, e^{-\mathcal{Q}(V)}  \nonumber \\
	&
	= \beta_a ({\rm Tr} (AB) {\rm Tr} (CD) + {\rm Tr} (AD){\rm Tr} (CB) ) + \beta_b( {\rm Tr} (ABCD) + {\rm Tr} (ADCB))\;. 
	\label{do}
\end{align}

\section{Double-winding Wilson loop and the monopole-only average}
\label{mon-only}

Here, we shall consider the Wilson loop average in a monopole-only Abelian-projected scenario (non-collimated monopole fluxes). The ensemble is formed by  closed worldlines with all possible adjoint charge distributions. For example, a single physical monopole loop $\gamma$ carrying charge $\alpha$ (positive roots of $\mathfrak{su}(N)$) is introduced by a source $\mathcal{J}_{\mu\nu}$, whose divergence is localized at the loop,
\begin{align}
\partial_\nu \mathcal{J}_{\mu\nu}= -2\pi 2N j_\mu \makebox[.5in]{,}j_\mu=\alpha\cdot T\int_\gamma \, ds \frac{d x_\mu}{ds}\delta(x-x(s))\;. 
\end{align}  
 The Wilson loop average for fundamental quarks reads
\begin{gather}  
\langle W_{\mathcal{C}_{\rm e}} \rangle   = \sum_{\{\gamma\} } \int[DA]\, \frac{1}{N} \, {\rm Tr}\,   \left(  e^{i \int_{\mathcal{C}} dx_\mu\,  A_{\mu}(x)  }  \right) e^{-\int d^4x \frac{1}{4g^2}(\tilde{F}_{\mu\nu}(A)-\mathcal{J}_{\mu\nu})^2}e^{-S(\{\gamma\})} \;,
\end{gather}
where $A_\mu$ is in the Cartan subalgebra, $\tilde{F}_{\mu\nu}$ is the dual field-strength, and $S(\{\gamma\})$ is the monopole action for a given distribution $\{\gamma\}$ of loops and adjoint weights. Performing the trace, we have
\begin{eqnarray}
 & & \langle W_{\mathcal{C}_{\rm e}} \rangle  =\frac{1}{N}\sum_i  I_{\omega_i}  \makebox[.5in]{,} I_{\omega}= \sum_{\{\gamma\}}\,  \int [DA] \, e^{i \int d^4x\, (\tilde{F}_{\mu \nu},\,\omega\cdot T s_{\mu \nu})  } e^{-\int d^4x \frac{1}{4g^2}(\tilde{F}_{\mu\nu}(A)-\mathcal{J}_{\mu\nu})^2}e^{-S(\{\gamma\})} \nonumber \;,
\end{eqnarray}
where $\omega_i$ are the weights of the defining representation.  The action for $A_\mu$ can be linearized with an auxiliary field $\lambda_{\mu\nu}$:
\begin{gather}
I_\omega =  \sum_{\{\gamma\}} \int[DA]\,[D\lambda]\,  e^{i \int d^4x\, (\tilde{F}_{\mu \nu} ,\, \omega\cdot T s_{\mu \nu} )}\, e^{\int d^4x\, \big(-\frac{g^2}{4}\lambda_{\mu\nu}^2+i(\lambda_{\mu\nu}, \tilde{F}_{\mu\nu}-\mathcal{J}_{\mu\nu})\big)}e^{-S(\{\gamma\})}\;.
\end{gather} 
 Next, the integration over $A_\mu$ leads to the constraint 
$
\epsilon_{\mu\nu\rho\sigma }\partial_\nu(\omega s_{\rho\sigma}+\lambda_{\rho\sigma})=0 $, 
which is solved by 
\begin{gather}
\lambda_{\mu\nu}=\partial_\mu\lambda_\nu-\partial_\nu \lambda_\mu-\omega s_{\mu\nu}\;.
\end{gather}
Then, after the field redefinition $2\pi 2N \lambda_\mu \to \lambda_\mu$, we get 
\begin{gather}
 I_\omega =  \int [D\lambda] \, e^{-\int d^4x\,  \frac{1}{\tilde{g}^2}(\partial_\mu\lambda_\nu-\partial_\nu\lambda_\mu-J_{\mu\nu})^2}\sum_{\{\gamma\}}  e^{i\int d^4x \,\big(\lambda_\mu,  j_\mu(\{\gamma\})\big)} e^{-S(\{\gamma\})}\;,
\end{gather}
 where $\beta = 2N \omega$, $J_{\mu\nu}= 2\pi \beta\cdot T\, s_{\mu\nu}(\mathcal{S}_{\rm e})$, and we defined 
\begin{gather}
\frac{1}{\tilde{g}^2}=\frac{g^2}{4} \frac{1}{16\pi^2 N^2}\;.
\end{gather}
The monopole currents can be modeled as we did for the collimated case, by assuming independent diluted gases labeled by $\alpha$. Besides a tension parameter,  monopole repulsive interactions must be introduced to stabilize the condensed phase. The techniques of Refs. \cite{fgreen,kleinert,diff2,diff3,mixed} can be used to obtain the effective representation
\begin{gather}
 I_\omega =     \int [D\lambda]\, [D\phi_\alpha]\, e^{-\int d^4x\, \frac{1}{\tilde{g}^2}(\partial_\mu \lambda_\nu-\partial_\nu\lambda_\mu-J_{\mu\nu})^2}e^{-\sum_{\alpha} \int d^4x\,   \left( D_\mu \bar{\phi}_\alpha D_\mu \phi_{\alpha}+\mathcal{U}(\phi)\right)}\;,
\end{gather}
where $\phi_\alpha$ are the effective monopole fields and $D_\mu =\partial_\mu-i \lambda_\mu \cdot \alpha $.  In this representation, the action has the same form as that for the collimated case in Eq.  \eqref{S-abe}. 
Note also that the independent path-integrals, when considering $\omega =\omega_i$, give the same result for every $i$. In particular, in the saddle-point evaluation of $I_{\omega_i}$, the dual gauge field has the form
$\lambda_\mu = a(x) \,\beta_i  \cdot T\, \partial_\mu\xi $. Thus, the set of charges $ \omega_i \cdot \alpha $ appearing in $D_\mu $ is always the same, while the potential is symmetric in $\alpha$. Then, what changes is the set of monopole fields that rotate when encircling the flux tube, but not the corresponding action.  That is, for quarks in the defining representation, we can simply write
\begin{equation}
 \langle W_{\mathcal{C}_{\rm e}} \rangle    = I_\omega \;.
 \label{monlya}
\end{equation}
 When compared with the average for collimated center-vortices and chains, the important difference for a double-winding Wilson loop $\mathcal{C}_{\rm e}$ is that in the monopole-only average \eqref{monlya} there is a single defining weight $\beta$ associated to the whole surface $S_{\rm e}$.

\section*{Acknowledgments} We would like to acknowledge 
D. Weingarten for useful discussions. 
This work was partially supported by the Conselho Nacional de Desenvolvimento Cient\'ifico e Tecnol\'ogico (CNPq), and the S\~ao Paulo Research Foundation (FAPESP), grant no.
2023/18483-0.

\end{document}